\documentclass[aip,reprint]{revtex4-1}
\usepackage{graphicx}% Include figure files
\usepackage{dcolumn}% Align table columns on decimal point
\usepackage{bm}% bold math
\usepackage{hyperref}
\usepackage[utf8]{inputenc}
\usepackage[T1]{fontenc}
\usepackage{mathptmx}
\usepackage{etoolbox}
\usepackage{xcolor}
\draft % marks overfull lines with a black rule on the right
\usepackage{float}
\usepackage{amssymb} % for symbols

\begin{document}

\title{Anti-thixotropic dynamics in attractive colloidal dispersions: a shear restructuring driven by elastic stresses}

%\title{Transient dynamics of attractive colloidal dispersion following step-down of shear rate} %Title of paper
%\title{Transient dynamics of attractive carbon black dispersions following a step-down of shear rate flow} 
%\title{Dynamic response of carbon black dispersions to a step-down in shear rate: a multi-method study}
%Unraveling Anti-Thixotropic Behavior in Carbon Black dispersions: Insights from Multimodal Rheological Analysis"
%Dynamic Structural Responses in Carbon Black dispersions under Step-Down Shear: A Rheo-USAXS and EIS Approach"
%Critical Shear Rate-Induced Transitions in Carbon Black dispersions: A Study of Anti-Thixotropy and Structural Reorganization"
%Anti-Thixotropic Dynamics in Attractive Colloidal dispersions in shear re-structuration driven by elastic stresses
%Linking Microstructure to Rheological Memory"
% repeat the \author .. \affiliation  etc. as needed
% \email, \thanks, \homepage, \altaffiliation all apply to the current author.
% Explanatory text should go in the []'s, 
% actual e-mail address or url should go in the {}'s for \email and \homepage.
% Please use the appropriate macro for the type of information

% \affiliation command applies to all authors since the last \affiliation command. 
% The \affiliation command should follow the other information.

\author{Julien Bauland}
\affiliation{Univ Lyon, Ens de Lyon, CNRS, Laboratoire de Physique, 69342 Lyon, France}

\author{Gauthier Legrand}
\affiliation{Univ Lyon, Ens de Lyon, CNRS, Laboratoire de Physique, 69342 Lyon, France}

%\author{Narayanan Theyencheri}
%\affiliation{ESRF - The European Synchrotron,  38043 Grenoble Cedex, France}

\author{Sébastien Manneville}
\affiliation{Univ Lyon, Ens de Lyon, CNRS, Laboratoire de Physique, 69342 Lyon, France}
\affiliation{Institut Universitaire de France (IUF)}

\author{Thibaut Divoux}
\affiliation{Univ Lyon, Ens de Lyon, CNRS, Laboratoire de Physique, 69342 Lyon, France}

\author{Arnaud Poulesquen}
\affiliation{ISEC, DPME, SEME, LFCM, Université of Montpellier, Marcoule, France}

\author{Thomas Gibaud}
%\email[]{Your e-mail address}
\email[]{Corresponding author, thomas.gibaud@ens-lyon.fr}
\affiliation{Univ Lyon, Ens de Lyon, CNRS, Laboratoire de Physique, 69342 Lyon, France}

\date{\today}

\begin{abstract}
Due to rich rheological properties, dispersions of attractive colloidal particles are ubiquitous in industries. Specifically, upon experiencing a sudden reduction in shear rate, these dispersions may exhibit transient behaviors such as thixotropy-where viscosity increases over time-and anti-thixotropy, characterized by an initial viscosity decrease before reaching a steady state. While thixotropy has been described as a competition between structure buildup and disruption, the mechanisms of anti-thixotropy remain poorly understood. Here, we investigate the anti-thixotropic dynamics of carbon black particles dispersed in oil-a system known for exhibiting anti-thixotropy-through flow step-down experiments. Using a multi-technique approach combining rheology with velocimetry and structural characterizations, we show that viscosity decrease results from a decrease in wall slip concomitant to shear-induced structural rearrangements, indicating a transition from a dynamical network of fractal clusters into a network of loosely connected dense agglomerates. Additionally, after a characteristic anti-thixotropic time $\tau$, a steady flow is reached. This time $\tau$ diverges with increasing shear rate at a critical value corresponding to a Mason number of one, indicating that anti-thixotropy occurs only when colloidal attraction outweighs viscous forces. More precisely, we show that the structural rearrangement underpinning the viscosity decrease is mediated by elastic stresses $\sigma_e$, such that $\tau \propto \sigma_e^{-3}$. Finally, on long time scales, the steady state is linked to a microstructure with nearly zero yield stress, indicating a loss of flow memory. These findings provide a mechanism for anti-thixotropy and suggest pathways for controlling viscosity and yield stress in attractive colloidal dispersions.
\end{abstract}
%\pacs{}% insert suggested PACS numbers in braces on next line

\maketitle %\maketitle must follow title, authors, abstract and \pacs

\section{Introduction}

Dispersions of attractive colloidal particles are non-Newtonian fluids commonly encountered in major industries, such as food and personal care products~\cite{Gibaud2012,Cao2020}, construction materials~\cite{Chougnet2008}, and environmental sciences~\cite{Baskaran2022,Morariu2022}. Weak physical interactions between particles, such as Van der Waals forces, result in an attractive interaction potential with a strength similar to thermal energy. Beyond a critical particle volume fraction, attractiv
e forces lead to the formation of a space-spanning network that imparts elastic, solid-like properties to the dispersions. At low particle volume fractions and strong attractions, such solids are referred to as colloidal gels~\cite{Trappe2004}, characterized by a network with fractal-like organization. 

Playing a key role in their functionality, colloidal gels exhibit dual behavior between a solid and a liquid, depending on the magnitude of shear forces relative to a critical threshold called the 'yield stress' $\sigma_y$~\cite{Coussot2014,Bonn2017,Nelson2017}. When the applied stress exceeds the yield stress, colloidal gels exhibit a shear-induced solid-to-liquid transition referred to as 'yielding'~\cite{Divoux2010,Divoux2012,Gibaud2010}. For 'reversible' gels, including carbon black gels~\cite{Trappe2000}, clays~\cite{pignon1997,burger2024} and depletion gels~\cite{Koumakis2015},  this transition is fully reversible and the gel-state reform under rest.  In such reversible gels, the versatility of their mechanical properties results from their dynamical microstructure, that can be reversibly modified by external stimulus. Consequently, a 'memory' of the flow can be encoded in the material, referring to its ability to retain structural or mechanical changes induced by external stimuli, even after the stimuli are removed~\cite{Divoux2024}. This ability arises because these materials exist in an out-of-equilibrium state with a complex energy landscape, scattered with multiple local minima.  While some minima are inaccessible under normal thermal energy ($k_BT$), external stimuli such as shear provide the energy required to drive the gels into these otherwise unreachable states~\cite{Dages2022b}. Memory is thus encoded by enabling transitions between configurations under the influence of external stimuli. Unlike colloidal glasses, whose dense packing restricts significant structural variation, the open structure of colloidal gels —due to their low volume fraction— allows for a wide range of distinct configurations. This structural diversity results in pronounced changes in mechanical properties, making memory effects particularly evident in colloidal gels. Notably, experimental studies have demonstrated the effect of shear on the rheological properties of attractive dispersions, showing that a 'rheological memory' of the flow history can be encoded under both continuous~\cite{Ovarlez2013a,Koumakis2015,Sudreau2022,Dages2022b} and oscillatory shear~\cite{Moghimi2017,Das2022}.
%'Rheological memory' is now fairly well understood at high shear rates but remains elusive at low shear rate. 

Irrespective of the nature of the flow, the effect of shear on the structure of attractive dispersions can be rationalized by the Mason number $Mn$~\cite{Mewis2009,Jamali2020}. This dimensionless number is defined as the ratio of the viscous drag force acting on particles ($F_\mathrm{visc}$) to the attractive forces between two particles ($F_\mathrm{attr}$), such that $Mn = \frac{F_\mathrm{visc}}{F_\mathrm{attr}}=\frac{6\pi \eta_f a^2 \dot{\gamma}}{U/\delta}$ with $a$ (m) the particle radius,  $\dot{\gamma}$ the shear rate, $\eta_f$ the viscosity of the background fluid, and $U$ (N.m) and $\delta$ (m) the depth and width of the attractive interaction potential, respectively. At high Mason numbers, i.e., $Mn \gg 1$, shear forces surpass attractive forces, causing particles to be fully dispersed and effectively erasing any previously encoded memory. In this limit, the system is fluid, and its viscosity is driven by hydrodynamic stresses $\sigma_h$. As the flow strength decreases and approaches  $Mn = 1$, shear forces become comparable to attractive forces, resulting in microstructural rearrangements within the dispersion through which memory is encoded. At very low Mason numbers $Mn \ll 1$, shear forces are negligible, and the system behaves as an elastic solid~\cite{Bauland2023,Varga2018}. In this limit, the dispersion behaves as a solid, whose elasticity is driven by elastic stresses $\sigma_e$. 

The interplay between shear forces and the microstructure at varying $Mn$ leads to a finite timescale $\tau$ for the dispersion to adapt to a given flow strength and to reach a steady state. This adaptation results in transient dynamics, particularly pronounced when changing the flow rate from high to low $Mn$, where the system's response slows significantly. In this regime, two distinct transient phenomena have been identified: 'thixotropy' and 'anti-thixotropy'.

On the one hand, thixotropy~\cite{Mewis2009,Larson2019} refers to a continuous increase in viscosity following a flow step-down, i.e., a quick transition from high to low shear rates. From a microstructural point of view, thixotropy results from a competition between the structure build-up and destruction under shear, which is well captured by structural kinetics models~\cite{Dullaert2006}. The thixotropic time $\tau_{thix}$ can be experimentally determined by so-called 'hysteresis loops', which involve applying a decreasing ramp of shear rate followed by an identical increasing ramp~\cite{Divoux2013}. Accordingly, the extent of thixotropy in a system can be determined from the Mnemosyne Number $My = \tau_{thix}\dot{\gamma}$, defined as the product of the thixotropic time and the imposed rate of deformation~\cite{Jamali2022}. 

On the other hand, 'anti-thixotropy' is defined as a viscosity decrease following a flow step-down. Sometimes referred to as 'rheopexy'~\cite{Ovarlez2013a,Narayanan2017} or 'negative thixotropy'~\cite{Larsen2024}, anti-thixotropy has been reported for several attractive systems, where applying moderate shear rates typically leads to a viscosity decrease of an order of magnitude over extended timescales. These long transient dynamics have been attributed to the formation of dense agglomerates of particles that reduce the effective volume fraction of the dispersion, in comparison with fractal-like structures~\cite{medalia1963,Essadik2021, Osuji2008, Narayanan2017, Ovarlez2013a, Hipp2019, Wang2022}. More specifically, for carbon black dispersion in oil, Wang et al.~\cite{Wang2022} have shown that the long-lasting viscosity decay is associated with anti-thixotropy rather than a viscoelastic relaxation. Their measurements, based on orthogonal superposition, suggest that the structural densification underlying anti-thixotropy proceeds through interpenetration of fractal clusters. In the work of Hipp et al.~\cite{Hipp2019}, also carried on carbon black dispersions, the viscosity decay was rather attributed to an apparent behavior and the sedimentation of dense aggregates. Irrespective of the microstructural scenario, anti-thixotropy was also associated with a significant reduction of the yield stress, offering a means to tune the rheological properties of colloidal dispersion at will~\cite{Ovarlez2013a,Jiang2022}.
%However, Hipp et al.~\cite{Hipp2019} have carried out Rheo-SANS experiments and established that the densification corresponds more to the coalescence of fractal clusters into large and dense  agglomerates. In~\cite{Hipp2019}, it remains unclear whether this densification is due to anti-thixotropy or sedimentation.  In~\cite{Ovarlez2013a,Jiang2022}, the authors demonstrate that, alongside the viscosity decrease, anti-thixotropy also results in a significant reduction of the yield stress, offering a means to tune the rheological properties of colloidal dispersion at will.

Yet, the specific microstructural scenario underlying the anti-thixotropic behavior in colloidal diseprsions, as well as the physical mechanisms involved, remain to be clarified. In particular, what defines the upper shear rate limit of the thixotropic regime? What determines the duration of the transient regime needed to achieve a stationary state? Additionally, the potential contribution of heterogeneous flow profiles, i.e., variations of the local shear rate in the sample such as wall slip and shear banding~\cite{Divoux2015, gibaud2008, Gibaud2016,Ewoldt}, must be investigated, as these phenomena are typically observed near the yield stress.

%To address these issues, we conduct flow step-down experiments —which is the quintessential method to characterize anti-thixotropy— on carbon black particles (CB) dispersed in oil. The behavior of such dispersions is well-documented~\cite{Gibaud2020, Richards2023, Trappe2000, koga2005, koga2008}, making CB dispersion a reference system, yet with practical applications including ink production~\cite{Liu2021}, construction materials~\cite{Li2006}, and semi-solid flow batteries~\cite{Liu2023}.
To address these issues, we conduct rheological experiments on attractive colloids, specifically carbon black particles (CB), dispersed in oil. The behavior of such dispersions is well-documented~\cite{Gibaud2020, Richards2023, Trappe2000}, making CB dispersion a reference system. These dispersions not only exhibit anti-thixotropy~\cite{medalia1963, Richards2023, Ovarlez2013a, Helal2016, Hipp2019, Wang2022a} but also demonstrate delayed yielding~\cite{Gibaud:2010, Grenard:2014}, fatigue~\cite{Gibaud:2016, Perge:2014}, rheo-conductive properties~\cite{Helal2016,Richards2017}, sensitivity to flow cessations~\cite{Dages2022b, Bouthier2022b}, and rheo-acoustic properties~\cite{Gibaud2020a, Dages2021}. Additionally, CB dispersions have practical applications in various fields, including rubber reinforcement~\cite{koga2005,koga2008}, ink production~\cite{Liu2021}, cement~\cite{Li2006}, and semi-solid flow batteries~\cite{Liu2023}.

% Using a multi-method approach, we conduct classical flow step-down experiments, an abrupt transition from a high shear rate to a low shear rate, which is the quintessential method to characterize anti-thixotropy.

Using a multi-method approach, we couple rheometric tests with ultrasonic speckle velocimetry (USV) to measure local flow profiles, allowing us to distinguish between local and bulk contributions. To probe the hierarchical structure of CB dispersions and their connectivity in real time, we combine rheo-Ultra Small Angle X-ray Scattering (USAXS) and rheo-Electric Impedance Spectroscopy (EIS) measurements. First, USV reveals that the decrease in stress upon flow step-down is preceded by a wall-slip regime, during which the material organizes into a dynamic fractal network. Second, as shear propagates through the geometry gap, the fractal microstructure continuously rearranges into dense, large agglomerates, as confirmed by USAXS. Additionally, EIS shows that the dense and large agglomerates form a loosely connected dynamical network. The characteristic time $\tau$ of the anti-thixotropic response increases exponentially with the applied shear rate, up to a critical shear rate interpreted as a Mason number equal to one, where clusters become effectively sticky relative to shear forces. Finally, using flow cessation experiments, we show that the restructuring of the fractal structure as well the duration of the transient regime relies on elastic stresses that deform the partially connected fractal network. Interestingly, the steady state associated with the anti-thixotropic response results in a single structuring of the dispersions with a vanishingly small yield stress, indicating a lack of flow memory upon reaching steady state. 
%%%%%%%%%%%%%%%%%%%%%%%%%%%%%%%%%%%%%%%%%%%%%%%%%%%%

\section{Material and methods}
\label{s:mm}

\subsection{Carbon black dispersions}
Following refs.~\cite{Dages2022b,Bauland2024}, CB particles (Vulcan\textsuperscript{\textregistered}PF, Cabot, density $d_{cb} = 2.26 \pm 0.03$) are dispersed in mineral oil (RTM17 Mineral Oil Rotational Viscometer Standard, Paragon Scientific, viscosity $\eta_f = 252.1~\rm mPa.s$ at $\rm T = 25^{\circ}$C, density $d_{oil} = 0.871$) at mass fractions $c_w$ ranging from 4 to 10 $\%$ ($w\backslash w$). The corresponding volume fractions of CB particles $\phi_{r_0}$ were calculated as $ \phi_{r_0} = c_w/[c_w + {d_{cb}}/{d_{oil}}(1-c_w)]$. After mixing, dispersions are sonicated for $2~\rm h$ in an ultrasonic bath (Ultrasonic cleaner, DK Sonic\textsuperscript{\textregistered}, United-Kingdom) to ensure that all particles are fully dispersed. We note that a mass fraction of $10~\rm \%$ $w\backslash w$ ($\phi_{r_0} =0.041$) constitutes an upper limit beyond which particles cannot be properly dispersed.
Based on TEM images and SAXS~\cite{Bauland2024}, Vulcan\textsuperscript{\textregistered}PF particle are composed are composed of nodules of radius $a\simeq 20$~nm that are fused together to form an aggregate of radius $r_0\simeq 85$~nm with a fractal dimension $d_{fr_0}\simeq 2.8$. Vulcan PF particles are, therefore, small and compact aggregates in the spectrum of CB particles manufactured~\cite{fernandez2017}.
We specifically chose Vulcan\textsuperscript{\textregistered} PF for its small particle size, which is well-suited for capturing large structures in SAXS. Additionally, we selected a relatively viscous mineral oil to slow sedimentation, ensuring that anti-thixotropic effects remain distinct from sedimentation effects (see \ref{sec:app}- Sedimentation). The attraction potential of the CB particle is estimated~\cite{Varga2019} to have a depth $U=30$~k$_B$T and a range $\delta=0.7$~nm.

\subsection{Rheology}
Experiments were carried out with two stress-controlled rheometers: ($i$) a Haake RS6000 (Thermo Scientific) equipped with a coaxial cylinder geometry composed of two concentric polycarbonate cylinders (inner diameter $20~\rm mm$, outer diameter $22~\rm mm$, and height 40~mm), and ($ii$) an MCR 302 (Anton Paar) equipped with a parallel-plate geometry (diameter $50~\rm mm$, gap $e = 0.5~\rm mm$). Additionally, stress jumps experiments were performed on a strain-controlled rheometer (ARES-G2, TA instruments) equipped with a cone-and-plate geometry (diameter $40~\rm mm$, cone angle 2~$^{\circ}$, truncation gap~$47~\rm \mu m$). 

All rheological measurements were performed at $T=25^{\circ}$C. After loading the CB dispersions, a preshear, $\dot{\gamma} = 500~\rm s^{-1}$ is first applied for $60~\rm s$ to erase any shear history and fully rejuvenate the sample. Then, a flow step-down is applied from $\dot{\gamma} = 500~\rm s^{-1}$ to $\dot{\gamma}_0 \in [0.2, \ 100]$ s$^{-1}$ while measuring the resulting stress response $\sigma(t)$. The stress-controlled rheometers successfully imposed $\dot{\gamma}_0$ in about $0.8~\rm s$ after the step flow, so data for $t \leq 0.8~\rm s$ are not considered (see Figure~\ref{fig:supFSD2} in Appendix). As a result, the thixotropic response of CB dispersions reported at short time scales~\cite{Wang2022} is not investigated. 

Finally, in addition to the flow step-down experiments, flow curves are performed by ramping down the shear rate from $\dot{\gamma} = $ 1000 to 0.01~$\rm s^{-1}$, using 10 points per decade with a duration $\Delta t = 1$, 50 and 100~$\rm s$ per point.

\subsection{Ultrasonic speckle velocimetry (USV)}

Velocity profiles are measured with ultrasound imaging using a homemade setup described in ref.~\cite{Gallot2013}. In brief, the local velocity $v(r,z,t)$ measured as a function of the position $r$ in the gap, the vertical position $z$ and time $t$. $v(r,z,t)$ is determined by cross-correlating successive ultrasonic speckle images of the dispersion. Hollow glass spheres with a mean diameter of  $6~\rm \mu m$ (SPHERICEL\textsuperscript{\textregistered} 110P8, Potters Industries) were added to the dispersion ($\phi = 0.9~\rm \% $) to act as acoustic contrast agents. Two-dimensional velocity maps were obtained using an array of 128 piezoelectric transducers arranged along the vertical direction over $32~\rm mm$, covering about two-thirds of the height of the coaxial cylinder geometry. The velocity profiles are identical at given time in the vertical direction and are therefore averaged. Velocity maps $v(r,t)$ were acquired every $6~\rm s$ for typically $2.10^3~\rm s$. To get statistically reliable data, the velocity maps were time-averaged over 100 successive cross-correlations.

%%%%%%%%%%%%%%%%%%%%%%%%%%%%%%%%%%%%%%%%%%%%%%%%%%%%%%%%%%%%%%%%%
\begin{figure}[t!]
    \includegraphics[scale=0.5, clip=true, trim=0mm 0mm 0mm 0mm]{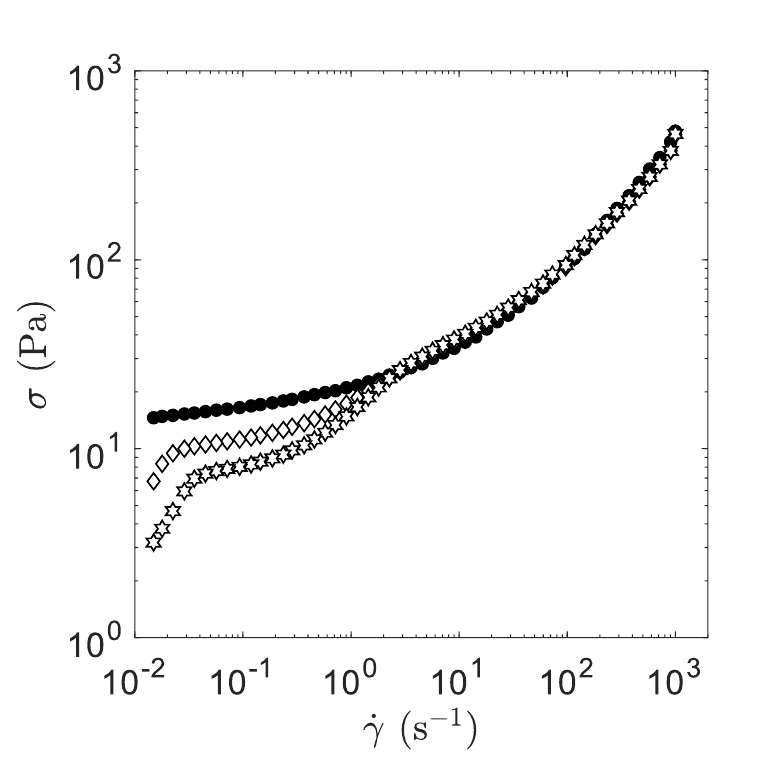}
    \centering
    \caption{Time-dependent apparent flow curves. Shear stress $\sigma$ vs shear rate $\dot{\gamma}$ during flow sweep experiment on 3.2~\% carbon black dispersion. Marker type codes for the shear duration at each shear rate: $\Delta t = 1$ ($\bullet$), 50 ($\lozenge$), and 100~s ($\star$).}
    \label{fig:suppFC}
\end{figure}
%%%%%%%%%%%%%%%%%%%%%%%%%%%%%%%%%%%%%%%%%%%%%%%%%%%%%%%%%%%%%%%%%

%%%%%%%%%%%%%%%%%%%%%%%%%%%%%%%%%%%%%%%%%%%%%%%%%%%%%%%%%%%%%%%%%
\begin{figure*}[th!]
    \includegraphics[scale=0.5, clip=true, trim=0mm 5mm 5mm 5mm]{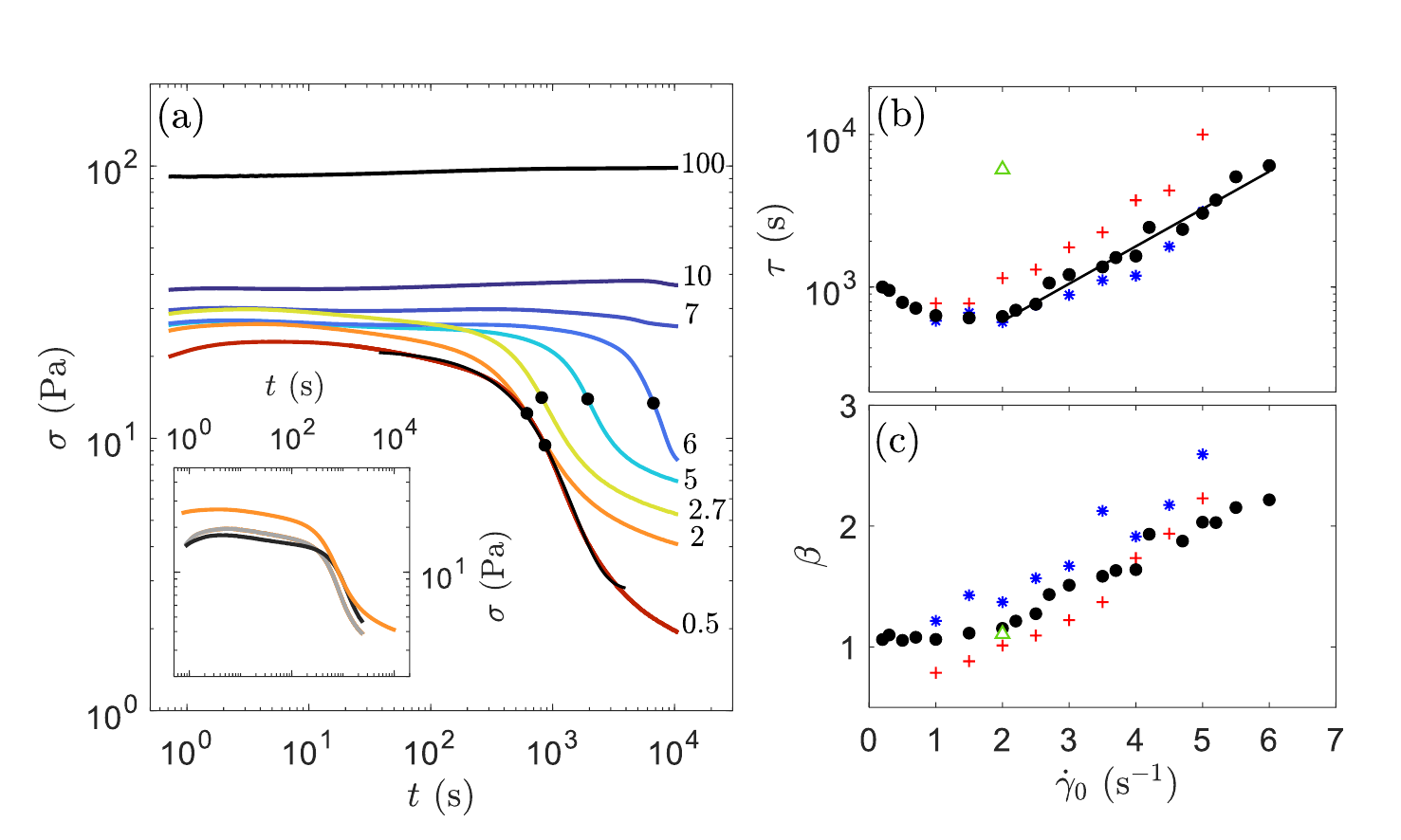}
    \centering
    \caption{Flow step-down experiments measured on the 3.2 \% carbon black dispersion. (a) Stress response $\sigma(t)$ following a flow step-down from $\dot{\gamma}=500~\rm s^{-1}$ to $\dot{\gamma}_0$. The values of $\dot{\gamma}_0$ are displayed on the right side. The black line is an example of fit by a compressed exponential decay $\sigma(t) = \sigma_0 + \sigma_1\exp[-(t / \tau)^{\beta}]$, at low shear rates $\dot{\gamma}_0 < 7~\rm s^{-1}$. Black dots indicate $t = \tau$, the characteristic time of the compressed exponential decay. Inset displays $\sigma$ vs.~time $t$ with $\dot{\gamma}_0=2~\rm s^{-1}$ in coaxial cylinder geometry (orange) and in parallel plate geometry with a gap size set at $0.25$ and $2~\rm mm$ (overlaid gray and black curves). (b) Characteristic time $\tau$ vs. shear rate $\dot{\gamma}_0$ for  volume fractions of CB: $\phi=4.1$ ($\ast$), 3.2 ($\bullet$), 2.4 ($+$) and 2 \% ($\vartriangle$). The black line is the best exponential fit for the 3.2 \% data, following $\tau = \tau_0\exp(\dot{\gamma}_0/\dot{\gamma}_1^*)$, with $\tau_0=193~\rm s$ and $\dot{\gamma}_1^* = 1.8~\rm s^{-1}$. (c) Exponent $\beta$ of the compressed exponential model vs. $\dot{\gamma}_0$. 
    }
    \label{fig:rheol}
\end{figure*}
%%%%%%%%%%%%%%%%%%%%%%%%%%%%%%%%%%%%%%%%%%%%%%%%%%%%%%%%%%%%%%%%%

\subsection{Ultra Small Angle X-ray Scattering (USAXS)}

The microstructural characteristics of the CB dispersion are probed using rheo-USAXS measurements conducted at the ID02 beamline within the European Synchrotron Radiation Facility (ESRF) in Grenoble, France~\cite{narayanan2020}. The incident X-ray beam, with a wavelength of 0.1~nm, is collimated to dimensions of 50~$\mu$m vertically and 100~$\mu$m horizontally. Utilizing an Eiger2 4M pixel array detector, two-dimensional scattering patterns are acquired. The subsequent data reduction process is detailed in~\cite{Panine2003,Narayanan2022}. The scattering intensity $I(q)$ as function of the scattering wave vector $q$ was derived by subtracting the two-dimensional scattering profiles of the CB dispersions and that of the mineral oil. 
Importantly, the resulting scattering intensity remained isotropic throughout this study and an azimuthal average was performed to obtain a one-dimensional spectrum $I(q)$. 
Measurements were conducted in both radial and tangential configurations, yielding similar results, showing the isotropic nature of the dispersions microstructure over the tested $q$ range. 

\subsection{Electrical Impedance Spectroscopy (EIS)}

Rheo-electrical measurements are performed using a stressed-controlled rheometer (MCR 302, Anton Par) equipped with a commercial set-up (Electro-Rheological Device, Anton Paar) consisting of a conductive $50~\rm mm$ parallel plate geometry. Electrical contact with the moving plate (rotor) is ensured by a liquid metal (Gallium-Indium eutectic $75.5 - 24.5 \%~\rm wt.$, ThermoFisher) to minimize solid friction~\cite{Helal2016}. A variable voltage of magnitude $10~\rm mV$ is applied at frequencies $f_e \in [2-5.10^4]~\rm Hz$ using a potentiometer (SP-300 Potentiostat, Biologic). Frequency sweep allows for the measurement of the complex electrical impedance, $Z^*(f_e) = Z^{\prime}(f_e) - iZ^{\prime\prime}(f_e)$, where $Z'$ and $Z''$ stands for the real and imaginary part of $Z^*$, every $30~\rm s$ during experiments lasting typically $2.10^3~\rm s$. A test at zero gap, i.e., when the two plates are in contact, shows that the set-up resistance is constant up to $f_e \leq 2.10^3~\rm Hz$, thus only data acquired in the range $f_e \in [2, \ 2.10^3]~\rm Hz$ are considered for analysis.

\section{Results}
\label{s:results}

\subsection{Time-dependent flow properties at low shear rate}

The CB dispersion at a volume fraction $\phi_{r_0}=3.2~\%$ displays a flow curve typical of yield stress fluids~\cite{Bauland2024}. As shown in Figure~\ref{fig:suppFC}, focusing on the instantaneous flow curve (1~s/points starting from high shear), we observe that, at high shear rates, the dispersion flows, and the stress decreases with decreasing shear rate, reaching a plateau at low shear rates. The limit at zero shear rate indicates the dynamic yield stress, $\sigma_y$, which is the minimum stress required to initiate or sustain flow in a material that has already been deformed or is in motion. By increasing the measurement duration per point from 1 to 100~s, we observe that the flow curve remains identical at high shear rates, which is not the case at low shear rates. Below $\dot{\gamma} \sim 1$~s$^{-1}$, or for stresses near $\sigma_y$~\cite{Hipp2019}, the apparent flow curve $\sigma(\dot \gamma)$ is time-dependent. At a given shear rate, the stress decreases when the measurement duration increases. This behavior is consistent with anti-thixotropy but could also result from wall-slip~\cite{Ewoldt,Divoux2015}. 
Furthermore, while flow curve measurements at varying rates offer a momentary view of the stress, they convolve time evolution with a continuous decrease of the shear rate. A more suitable protocol consists in conducting flow step-down experiments. These experiments involve switching from a high shear rate to a low shear rate as fast as possible, while monitoring the stress evolution over longer time. This experiment highlights the response of the material to a sudden change in shear rate, providing deeper and clearer insights into its dynamics.

\subsection{Anti-thixotropic stress response during a flow step-down}
%%%%%%%%%%%%%%%%%%%%%%%%%%%%%%%%%%%%%%%%%%%%%%%%%%%%%%%%%%%%%%%%%
\begin{figure*}[t!]
    \includegraphics[scale=0.55, clip=true, trim=0mm 0mm 0mm 0mm]{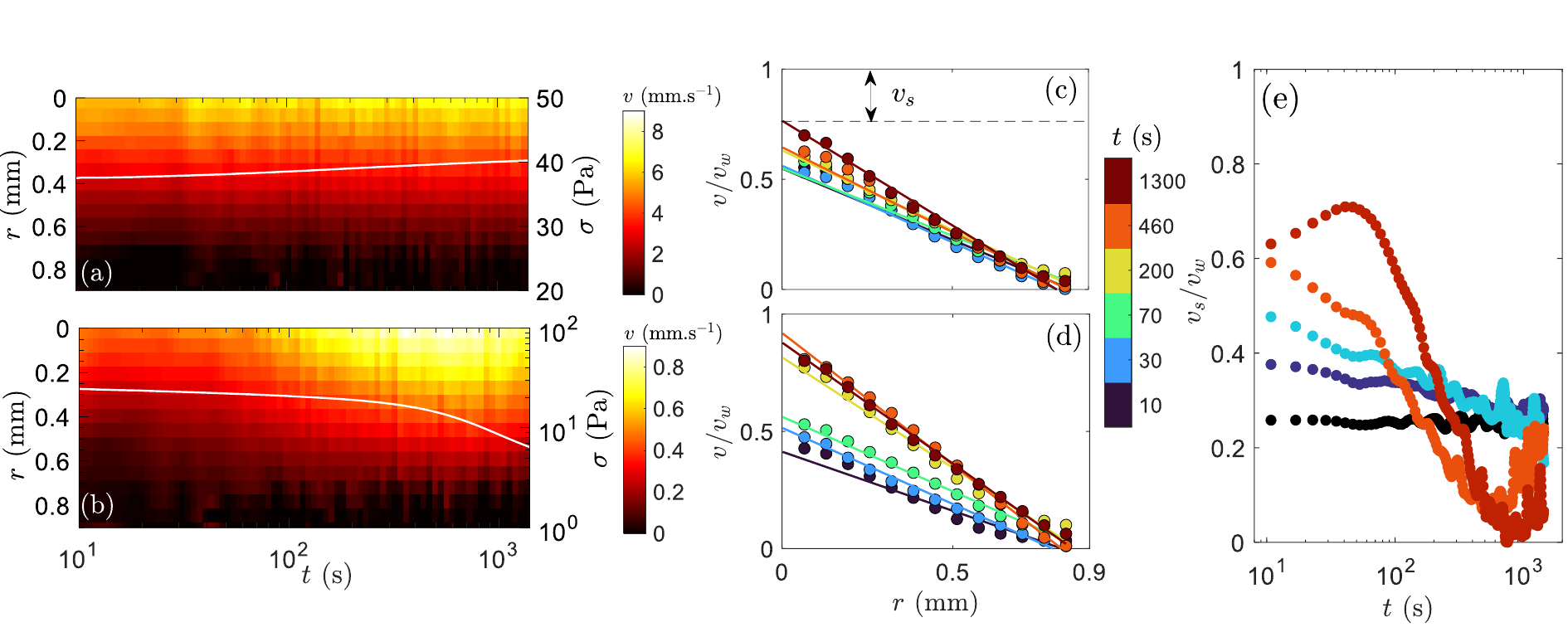}
    \centering
    \caption{Local velocity profiles measured on the 3.2 \% carbon black dispersion using ultrasonic speckle velocimetry (USV) during flow step-down. (a)-(b) Spatiotemporal diagrams of the velocity data $v(r,t)$ as a function of position $r$ and time $t$ for (a) $\dot{\gamma}_0 = 10~\rm s^{-1}$, and (b) $1~\rm s^{-1}$. The stress response $\sigma(t)$ is displayed as a white line (right axis). (c)-(d) Corresponding velocity profiles measured at different times after imposing flow step-down of $\dot{\gamma}_0 = 10$ (c) and $1~\rm s^{-1}$ (d). The fluid velocity $v$ normalized by the wall velocity $v_w$ is displayed as function of the the radial distance in the gap $r$. Solid lines are the best linear fit of the data. (e) Time dependence of the wall slip velocity $v_s$, defined as the velocity difference between the rotor wall and the fluid $v_s = v_w - v_f$, normalized by the wall velocity. Colors code for the applied shear rate $\dot{\gamma}_0 = 100$ (black), 10 (blue), 5 (cyan), 1 (orange), and $0.5~\rm s^{-1}$ (red). 
    }
    \label{fig:USV}
\end{figure*}
%%%%%%%%%%%%%%%%%%%%%%%%%%%%%%%%%%%%%%%%%%%%%%%%%%%%%%%%%%%%%%%%%
We now focus on the bulk rheology of CB dispersions under flow step-down, i.e., from $\dot{\gamma} = 500~\rm s^{-1}$ down to a shear rate of interest $\dot{\gamma}_0$. Figure~\ref{fig:rheol} shows the temporal evolution of the stress for the $3.2~\%$ CB dispersion quenched to different shear rates $\dot{\gamma}_0 \in [100-0.5]~\rm s^{-1}$. For $\dot{\gamma}_0 \geq 7~\rm s^{-1}$, we observe that a steady state is reached almost immediately, and the stress decreases by less than $10 \%$ over $10^4~\rm s$. In contrast, for $\dot{\gamma}_0 < 7~\rm s^{-1}$, the stress shows a delayed relaxation, i.e., it remains constant on short time scales and then decrease by a factor of 5 to 10 depending on the shear rate.  
The shape of the stress drop and the timescale over which it occurs are consistent with the anti-thixotropic response of CB dispersions, as previously reported~\cite{Ovarlez2013a,Wang2022,Hipp2019}. It highlights the robustness of this behavior across different CB particles and apolar solvents. Additionally, this phenomenology is consistently observed in parallel plate geometry, and for various gap sizes (see inset in Fig.~\ref{fig:rheol}), ruling out any confinement or finite-size effect. 

Our results point to a critical shear rate $\dot{\gamma}_0^* \approx 7~\rm s^{-1}$, which is robustly observed for various volume fractions of CB particles, $\phi_{r_0} \in [2.4,4.1]~\%$ (see Fig.~\ref{fig:supFSD} in Appendix). Note the value of $\dot{\gamma}_0^*$ is consistent with the value previously reported for CB particles in mineral oil~\cite{Wang2022}. Yet, $\dot{\gamma}_0^*$ depends on the nature of the solvent: a larger value of $\dot{\gamma}_0^*=200~\rm s^{-1}$ was found when CB particles dispersed in propylene carbonate~\cite{Hipp2019}. 

More quantitatively, the stress response $\sigma(t)$ associated with the anti-thixotropic behavior of the CB dispersions is well captured by a stretched exponential using the following expression: $\sigma = \sigma_0 + \sigma_1\exp[-(t/\tau)^{\beta}]$, where $\sigma_0$ denotes the stress value when the shear rate is decreased, $\sigma_1$ is a stress scale, $\tau$ is a characteristic time, and $\beta$ is an exponent controlling the shape of the exponential relaxation [see fit in Fig.~\ref{fig:rheol}(a)]. As shown in Figure~\ref{fig:rheol}(b), the time $\tau$ is roughly constant for $\dot{\gamma}_0 < 2~\rm s^{-1}$, and increases exponentially with the applied shear rate for $2 \leq \dot{\gamma}_0 \leq 6~\rm s^{-1}$.  This dependence contrasts with the behavior of thixotropic timescales, which typically decrease with increasing shear rate~\cite{Dullaert2006}. It also contrasts with the formation of log-rolling structures under confinement, where the time necessary to reach a steady state scales linearly with the applied shear rate~\cite{Varga2019,Grenard2011}. 
Moreover, while the critical shear rate $\dot{\gamma}_0^*$ appears independent of the volume fraction of primary CB particles $\phi_{r_0}$ within the tested range, the anti-thixotropic time $\tau$ increases with a decrease of $\phi_{r_0}$ for a given shear rate [Fig.~\ref{fig:rheol}(b)]. Imposing $\dot{\gamma}_0 = 2~\rm s^{-1}$ yields $\tau = 640~\rm s$ for $\phi_{r_0}= 3.2~\%$, and $\tau$ increases to $1140~\rm s$ and $5900~\rm s$ for $\phi_{r_0}= 2.4$ and $2~\%$, respectively.
Such an increase of $\tau$ with decreasing $\phi_{r_0}$ has also been reported for dispersions of latex particles~\cite{Mills1991a}. The exponent $\beta$ is positively correlated with $\tau$ and the exponential decay becomes more compressed ($\beta > 1$) when $\dot{\gamma}_0$ increases, yet without clear trend as a function of $\phi$ [Fig.~\ref{fig:rheol}(c)].

So far, we have shown that following a flow step-down, CB dispersions exhibit an anti-thixotropic drop in stress below a critical shear rate $\dot{\gamma}_0^* \approx 7~\rm s^{-1}$ that is independent of the tested volume fractions. The anti-thixotropic time $\tau$ increases exponentially with the applied shear rate, and also increases with the volume fraction. In the following sections, we focus on the $3.2~\%$ CB dispersions and couple rheometric measurements to USV, USAXS, and EIS measurements to provide further insights on the microstructural scenario underpinning this macroscopic rheological response. 

\subsection{Flow velocity profiles}

To investigate the potential contribution of heterogeneous flow to the macroscopic stress response evidenced in the previous section, we use USV to measure the local flow profile $v(r,t)$ a function of of the position within the gap of the rheometer $r$ and the time $t$ just after a step flow-down over typically $2.10^3~\rm s$. 
At a shear rate of $\dot{\gamma}_0=10~\mathrm{s}^{-1}$, both the stress and the flow profile are in a steady state [Fig.~\ref{fig:USV}(b)]. In contrast, at a shear rate of $\dot{\gamma}_0=1~\mathrm{s}^{-1}$ where anti-thixotropy is observed, both the stress and the flow profile evolve over time [Fig.~\ref{fig:USV}(c)].
Figure~\ref{fig:USV}(c)-(d) display typical examples of a velocity profiles, where the fluid velocity is plotted as a function of the radial distance $r$ in the gap, measured at $\dot{\gamma}_0=10$ and $1~\mathrm{s}^{-1}$, respectively. All velocity profiles remain linear, i.e., no shear banding is observed. However, significant wall slip is observed at the rotor. The fluid velocity at the wall $v_f$ is determined by a linear regression, and the degree of wall slip is then quantified by calculating the slip velocity defined as \cite{Divoux2015} $v_s = v_w - v_f $, where $v_w$ is the (inner) wall velocity.

%%%%%%%%%%%%%%%%%%%%%%%%%%%%%%%%%%%%%%%%%%%%%%%%%%%%%%%%%%%%%%%%%
\begin{figure*}[t!]
    \includegraphics[scale=0.52, clip=true, trim=0mm 0mm 0mm 0mm]{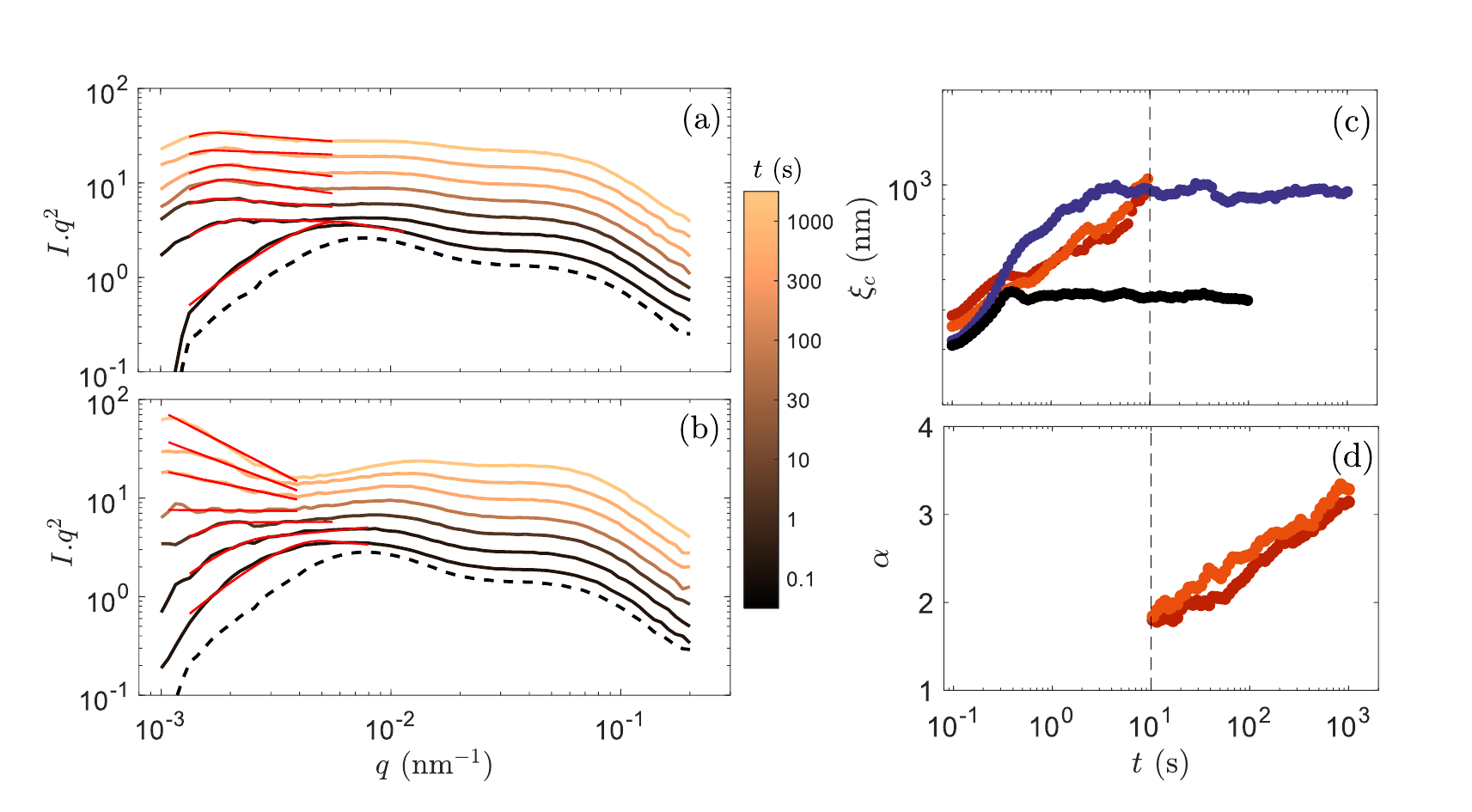}
    \centering
    \caption{Ultra-small angle X-ray scattering (USAXS) measurements conducted on the 3.2 \% of carbon black (CB) dispersion during flow step-down. (a)-(b) Kratky plots $I.q^2$ vs.~$q$ of the scattering profiles acquired at different times after imposing a shear rate $\dot{\gamma}_0 = 10~\rm s^{-1}$ (a) and $1~\rm s^{-1}$ (b). Black dotted curves are the scattering profile of the dispersion during the preshear step at $\dot{\gamma} = 500~\rm s^{-1}$. In (a)-(b), data are shifted along the $y$-axis for better visualization. (c)-(d) Quantitative analysis of the low-q region ($10^{-3} \leq q \leq 10^{-2}~\rm nm^{-1}$) using either a Guinier-Porod or a simple power law model. The radius $\xi_c$ of the Guinier–Porod or the power-law exponent $\alpha$ are shown as a function of time in panels (c) and (d), respectively, depending on which model best fits the data. Colors code for the applied shear rate $\dot{\gamma}_0 = 100$ (black), 10 (blue), 1 (orange) and $0.5~\rm s^{-1}$ (red). Examples of fit are displayed as red lines in (a)-(b) panels.
    }
    \label{fig:SAXS}
\end{figure*}
%%%%%%%%%%%%%%%%%%%%%%%%%%%%%%%%%%%%%%%%%%%%%%%%%%%%%%%%%%%%%%%%%

The normalized slip velocity $v_s/v_w$ is displayed in Figure~\ref{fig:USV}(e) against the time elapsed since the flow step-down. For all applied shear rates, the velocity profiles display some degree of slippage. For $\dot{\gamma}_0=100~\rm s^{-1}$, previously associated with a quasi-instantaneous steady-state based on the stress response, $v_s / v_w$ is about $30~\%$. Second, the initial slip velocity measured at $t\approx 10~\rm s$ decreases for increasing $\dot{\gamma}_0$. For $\dot{\gamma}_0 = 1$ and $0.5~\rm s^{-1}$, $v_s / v_w$ is as high as $60~\%$ shortly after imposing $\dot{\gamma}_0$. For $\dot{\gamma}_0=100~\rm s^{-1}$, $v_s / v_w$ is constant while at lower shear rates, $v_s / v_w$ decreases over time and tends to about $30~\%$. For $\dot{\gamma}_0 = 0.5$ and $1~\rm s^{-1}$, $v_s / v_w$ shows a non-monotonic evolution and goes through a minimum where $v_s=0$, showing an absence of slip. The time corresponding to $v_s=0$ nicely matches the anti-thixotropic time, with $\tau = 650$ and $850~\rm s$ for $\dot{\gamma}_0 = 0.5$ and $1~\rm s^{-1}$, respectively. The correlation between the macroscopic stress decay and local velocity profiles at $\dot{\gamma}_0 = 1~\rm s^{-1}$ is further illustrated on the spatiotemporal diagram in Figure~\ref{fig:USV}(b). The decrease of the stress $\sigma$  starts when the fluid velocity approaches the wall velocity. 

In summary, the slip velocity strongly correlates with the evolution of the shear stress. At low shear rates, specifically $\dot{\gamma}_0 = 0.5$ and $1~\rm s^{-1}$, the simultaneous decrease of the slip velocity $v_s$ and the stress $\sigma$ suggests that the material undergoes some structural reorganization under shear. In the following section, we investigate the microstructure of the CB dispersion under flow using USAXS and EIS.

\subsection{Microstructure under shear}

\subsubsection{Rheo-X-ray scattering measurements}

Rheo-USAXS experiments are performed to probe the microstructure of the $3.2~\%$ CB dispersion over length scales ranging between $30~\rm nm$ and $3~\rm \mu m$. We measure the 2D scattered intensity $I(q)$ as a function of the scattering wave vector $q$ during flow step-down experiments. At these tested length scales, the scattering intensity $I(q)$ remains isotropic, with no sedimentation detected over the time scales examined (see Fig.~\ref{fig:supSedim} in Appendix).

Figure~\ref{fig:SAXS}(a) displays Kratky plots of the 1D scattering intensity, i.e., $I.q^2$ vs.~$q$, at various times under shear at $\dot{\gamma}_0 = 10~\rm s^{-1}$. The spectra reveal two to three structural length scales, depending on the shear rate, in agreement with previous works~\cite{Dages2022b,Bauland2024}. During the rejuvenation step at 500 s$^{-1}$ [indicated by black dotted lines in Fig.~\ref{fig:SAXS}(a)], the structure of the dispersion shows two distinct peaks centered around $q \approx 7.10^{-2}$ and $q \approx 8.10^{-3}~\rm nm^{-1}$. These peaks correspond to the form factor of the CB particles of size $r_0$ and to the structuring of the CB particles into small fractal clusters of size $\xi_{c_1}$ and fractal dimension $d_{f_1}$, respectively. After the flow step down, we observe the emergence of three length scales as observed by Koga et al. \cite{koga2005, koga2008} when CB particles are dispersed in a rubber matrix. 
This corresponds to the CB particles that  aggregate into small clusters of size $\xi_{c_1}$ and fractal dimension $d_{f_1}$ inherited from the rejuvenation step that further aggregate into large clusters of size $\xi_c$ and fractal dimension $d_f$.

When quenching to high shear rate such as $\dot{\gamma}_0 = 10~\rm s^{-1}$, the large cluster size quickly grows and then stabilizes at $q \approx 2.10^{-3}~\rm nm^{-1}$ with no further changes observed in the scattering spectrum up to $t = 10^{3}~\rm s$. The same microstructure can be reproduced using a different shear protocol, namely a flow sweep, (see Fig.\ref{fig:supSAXS2} in Appendix), demonstrating that for $\dot{\gamma} \geq 10~\rm s^{-1}$, the microstructure of the dispersion is unaffected by the flow history.  
In contrast, in the anti-thixotropic regime, at $\dot{\gamma} < 10~\rm s^{-1}$, the scenario is different. Indeed, the Kratky plot of the scattered intensity obtained when imposing $\dot{\gamma}_0 = 1~\rm s^{-1}$ shows a gradual increase of the large cluster size until $t \approx 30~\rm s$ [Fig.~\ref{fig:SAXS}(b)]. Beyond this time, the low-$q$ intensity scales as a power law $I \propto q^{-\alpha}$, indicating that the cluster size exceeds the detection range of the USAXS setup.

%%%%%%%%%%%%%%%%%%%%%%%%%%%%%%%%%%%%%%%%%%%%%%%%%%%%%%%%%%%%%%%%%%%%%%%%%%%%%%%%%%%
\begin{figure*}[!t]
    \includegraphics[scale=0.55, clip=true, trim=3mm 0mm 0mm 0mm]{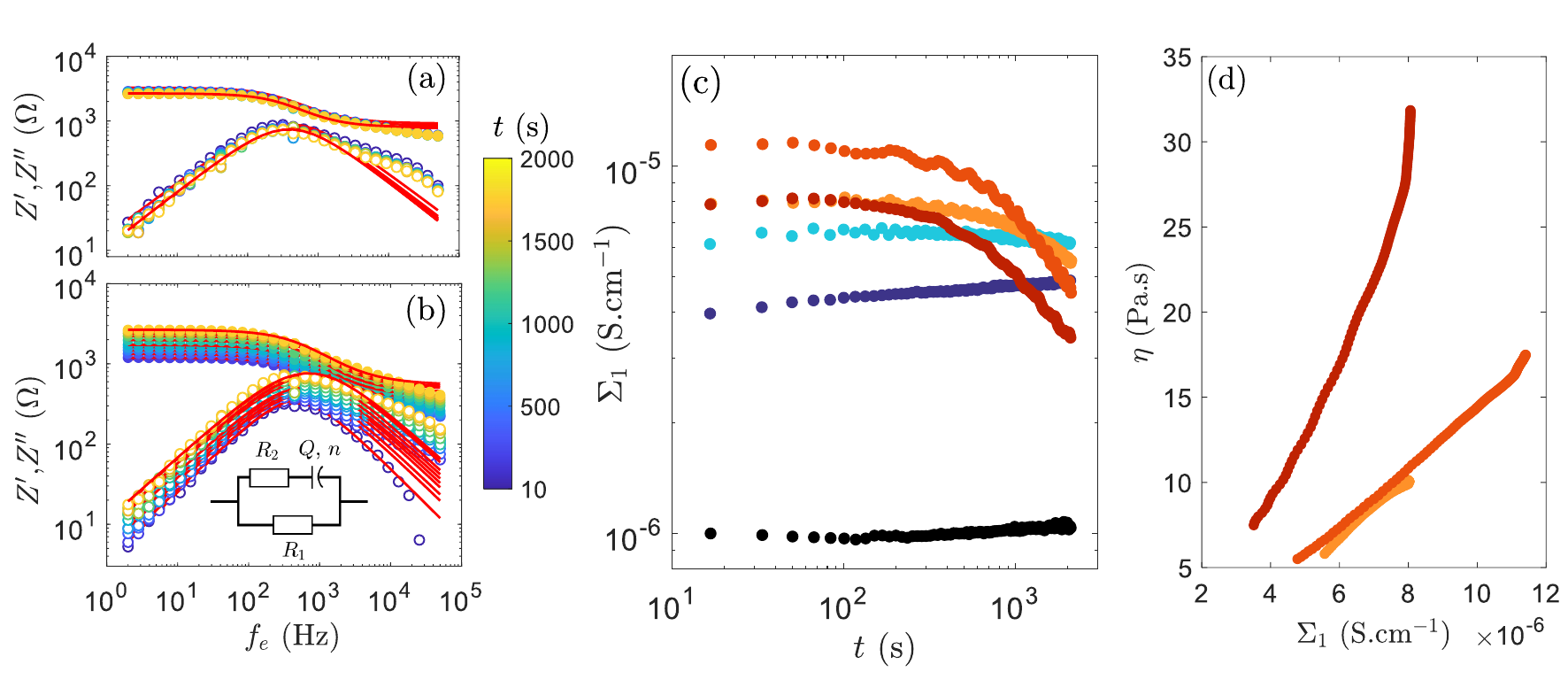}
    \centering
    \caption{Electrical impedance spectroscopy (EIS) measurements on a 3.2 \% of carbon black (CB) dispersion during flow step-down. (a)-(b) Real part ($Z^{\prime}$) and imaginary part ($Z^{\prime\prime}$) of the complex impedance as a function of the input voltage frequency $f_e$ measured at different times after stepping down the shear rate from $500~\rm s^{-1}$ to $\dot{\gamma}_0 = 10~\rm s^{-1}$ (a) and $1~\rm s^{-1}$ (b). Red curves are the best fits of the electrical model depicted in the inset of (b). (c) Temporal evolution of the conductivity $\Sigma_1$ associated with the structuring of carbon black particles. Colors code for the applied shear rate $\dot{\gamma}_0 = 100$ (black), 10 (blue), 5 (cyan), 2 (light orange), 1 (dark orange) and $0.5~\rm s^{-1}$ (red). On the right panel, the same conductivity $\Sigma_1$ is displayed as a function of the time elapsed during the flow step-down. (d) Viscosity $\eta$ vs.~$\Sigma_1$ with identical color code.
    }
    \label{fig:elec}
\end{figure*}
%%%%%%%%%%%%%%%%%%%%%%%%%%%%%%%%%%%%%%%%%%%%%%%%%%%%%%%%%%%%%%%%%%%%%%%%%%%%%%%%%%%%

We now focus on the large-scale structures (sizes superior or equal to $\xi_c$) that are most affected in the anti-thixotropic regime.
To capture the dynamics of structural changes, we use either a Guinier-Porod or a simple power-law model within the range $1.10^{-3} \leq q \leq 1.10^{-2}~\rm nm^{-1}$, depending on which model best fits the data [see red curves in Fig.~\ref{fig:SAXS}(a)-(b)]. The fitting parameters —the radius $\xi_c$ for the Guinier-Porod and the power-law exponents $\alpha$— are reported in Figure~\ref{fig:SAXS}(c)-(d). For $\dot{\gamma}_0 = 0.5$ and $1~\rm s^{-1}$, the transition between the two models is highlighted by a dotted vertical line. At higher shear rates, i.e., $\dot{\gamma}_0 = 100$ and $10~\rm s^{-1}$, $\xi_c$ increases for $t \approx 0.5$ and $3~\rm s$, respectively, and then remains constant as shown in Figure~\ref{fig:SAXS}(a). It indicates a quick transition toward a steady state as hypothesized for such high shear rates. In contrast, for $\dot{\gamma}_0 = 0.5$ and $1~\rm s^{-1}$, $\xi_c$ increases for $t \approx 10~\rm s$. From this time, data are best described by the power law model, with an exponent $\alpha = 2$, indicating the formation of a fractal-like network with a fractal dimension $d_f \approx 2$, and a mesh size beyond the USAXS detection range. Following this, the exponent $\alpha$ increases to $3.2$, reflecting a continuous rearrangement of the network into a denser microstructure. 

The power-law increase in intensity, with $\alpha > 3$, indicates that scattering is dominated by rough interfaces between the CB phases and the solvent. In this case, the fractal dimension of the interface $d_{fs}$ can be determined using the relation $\alpha = 6-d_{fs}$~\cite{Beaucage1994}. Comparable exponent values have been observed in similar contexts, such as the anti-thixotropic response of CB particles in low viscosity solvent ($\alpha = 3.6$)~\cite{Hipp2019}, and CB dispersions exposed to high power ultra-sound ($\alpha = 3.7$)~\cite{Dages2021}. These rough interfaces can be attributed to (i) micro-cracks forming within the CB network or (ii) the formation of large, dense agglomerates of CB particles. Supporting the second explanation, electron microscopy images of sheared polystyrene latex particle dispersions, which also display anti-thixotropy, show the formation of large agglomerates composed of densely packed particles~\cite{Mills1991a}. It suggests that the formation of dense agglomerates in CB dispersions is more likely, which also aligns with previous hypothesis regarding the structure of CB dispersions, as determined by dielectric measurements~\cite{Narayanan2017} and orthogonal superposition rheometry~\cite{Wang2022}. 

Finally, in Figure~\ref{fig:SAXS}(c)-(d), the gradual evolution of $\xi_c$ and $\alpha$ reflects a continuous structural building and rearrangement within the sample during shear. This is consistent with USV measurements, which confirms that the dispersion does not exhibit plug flow, even at short times. For shear rates $\dot{\gamma}_0 = 0.5$ and $1~\rm s^{-1}$, the anti-thixotropic time $\tau$ roughly corresponds to the time when $\alpha = 3$, 
indicating that anti-thixotropy  corresponds to a gradual restructuring of the dispersion, transitioning from fractal clusters to large, dense agglomerates. Considering the delay before the onset of anti-thixotropic stress decay, structural changes appear to have little effect on the measured stress or may be counterbalanced by the decrease in $v_s$, until $t \approx \tau$. 

In conclusion, the steady-state flow at shear rates of $\dot{\gamma} = 100~\rm s^{-1}$ and $10~\rm s^{-1}$ is characterized by a rapid evolution of the large cluster size. In contrast, the long-lasting transient regime observed during the anti-thixotropic response at $\dot{\gamma} = 0.5$ and $1~\rm s^{-1}$ arises from the gradual transition from large fractal clusters into large and dense agglomerates which size exceed the range accessible using USAXS. 

\subsubsection{Rheo-Electrical impedance measurements}

To further characterize the microstructure of the dispersion, we take advantage of the conductive properties of CB particles. We perform EIS measurements under shear by applying an alternating voltage of frequency $f_e$, and determining the complex impedance $Z^*(f_e)$ from the measured current. Figures~\ref{fig:elec}(a)-(b) display the real $Z^{\prime}(f_e)$ and imaginary $Z^{\prime\prime}(f_e)$ parts of $Z^*$ measured at different times after imposing $\dot{\gamma}_0 = 10~\rm s^{-1}$ and $1~\rm s^{-1}$, respectively. The shape of $Z^{\prime}(f_e)$ and $Z^{\prime\prime}(f_e)$ can be described by an electrical model composed of a resistance $R_1$ in parallel with a second resistance $R_2$ and a constant phase element of parameters $Q$ and $n$~\cite{Cole1928} [see inset in Fig.~\ref{fig:elec}(b)]. In this model, $R_1$ is associated with the CB particles \cite{Legrand2022}, allowing us to calculate $\Sigma_1$, the conductivity associated with the percolated network of CB particles (see Appendix for more details). 

Figure~\ref{fig:elec}(c) shows the temporal evolution of $\Sigma_1$ after applying $\dot{\gamma}_0$ = 0.5, 1, 2, 5, 10, and 100$~\rm s^{-1}$. At early times, i.e., $t<100~\rm s$, the conductivity increases as the shear rate decreases, indicating a more connected microstructure at lower shear rates. For $\dot{\gamma}_0 = 100$ and $10~\rm s^{-1}$, the conductivity quickly reaches a quasi steady-state, indicating a rapid transition to equilibrium, as also seen in the rheometric and USAXS data. Specifically, for $\dot{\gamma}_0 = 100~\rm s^{-1}$, the low conductivity of the dispersion is consistent with isolated clusters within the dispersion. At low shear rates, namely $\dot{\gamma}_0 =$ 0.5, 1 and 2$~\rm s^{-1}$, the conductivity decreases with shearing time, corresponding to the decrease in stress measured by rheometry. This decrease in conductivity suggests the formation of denser but less connected agglomerates~\cite{Narayanan2017, Richards2017}. For $\dot{\gamma}_0 = 5~\rm s^{-1}$, the conductivity only shows an initial decrease, when the maximum shear time measured by EIS is shorter than the anti-thixotropic time, $\tau = 3250~\rm s$. In Figure~\ref{fig:elec}(d),  the shear viscosity for $\dot{\gamma}_0 = 1$ and $2~\rm s^{-1}$, increases linearly with the conductivity, indicating that the structural rearrangements probed by dielectric measurements directly account for the evolution of the rheological properties. However, for $\dot{\gamma}_0 = 0.5~\rm s^{-1}$, the relationship $\Sigma_1 (\eta)$ is non-linear, possibly due to additional slip effects in the parallel-plate geometry. 

In conclusion, consistent with USAXS data, EIS measurements confirm that anti-thixotropy arises from the reorganization of a homogeneous, well-connected structure into weakly connected and heterogeneous structures. 

\section{Discussion}

\subsection{Critical Mason number}

We have observed anti-thixotropy in a CB dispersion at shear rates below a critical value $\dot{\gamma}_0^* \approx 7~\rm s^{-1}$. This phenomenology is robustly observed across the following volume fractions in CB particles: $\phi_{r_0} =$ 4.1, 3.2, and 2.4~\%. Similar critical shear rates have been reported in the literature and interpreted using the inverse of a Bingham number, $Bi^{-1} = \sigma / \sigma_y \simeq 1$, meaning that the instantaneous stress $\sigma$ measured upon applying $\dot{\gamma}_0$ is close to the dynamic yield stress $\sigma_y$~\cite{Hipp2019}. Applying this concept to our results, for a dispersion at $\phi_{r_0} = 3.2~\%$, the instantaneous stress measured before the onset of  the anti-thixotropic response ranges between 20 and 30 Pa (Fig.~\ref{fig:rheol}), which approximately corresponds to a $Bi^{-1} \simeq 1$ when considering $\sigma_y = 14.7~\rm Pa$, as measured in the fast flow curve (Fig.~\ref{fig:suppFC}). 

To bring more insights to the physical understanding of $\dot{\gamma}_0^*$, we instead refer to the Mason number, $Mn=\frac{6\pi \eta_f a^2 \dot{\gamma}}{U/\delta}$. For clusters of particles, the relevant length scale experiencing the drag force is not the CB particle itself but rather the cluster size $a = \xi_c$, estimated to be about $1.4~\rm \mu m$ for the $3.2~\%$ dispersion at low shear rate~\cite{Bauland2024}. Assuming $\delta = 0.7~\rm nm$ and $U$ comprised between $20$ and $30~\rm k_BT$~\cite{Varga2019}, the critical Mason number Mn$^* \sim 1$ is reached for $\dot{\gamma} \in [11-20]~\rm s^{-1}$, in fair agreement with the experimental value $\dot{\gamma}_0^* \approx 7~\rm s^{-1}$. Thus, $\dot{\gamma}_0^*$ is interpreted as a threshold where inter-cluster attraction forces are of the order of the shear forces. 

%%%%%%%%%%%%%%%%%%%%%%%%%%%%%%%%%%%%%%%%%%%%%%%%%%%%%%%%%%%%%%%%%%%%%%%%%%%%%%%%%%
\begin{figure}
    \includegraphics[scale=0.5, clip=true, trim=0mm 0mm 0mm 0mm]{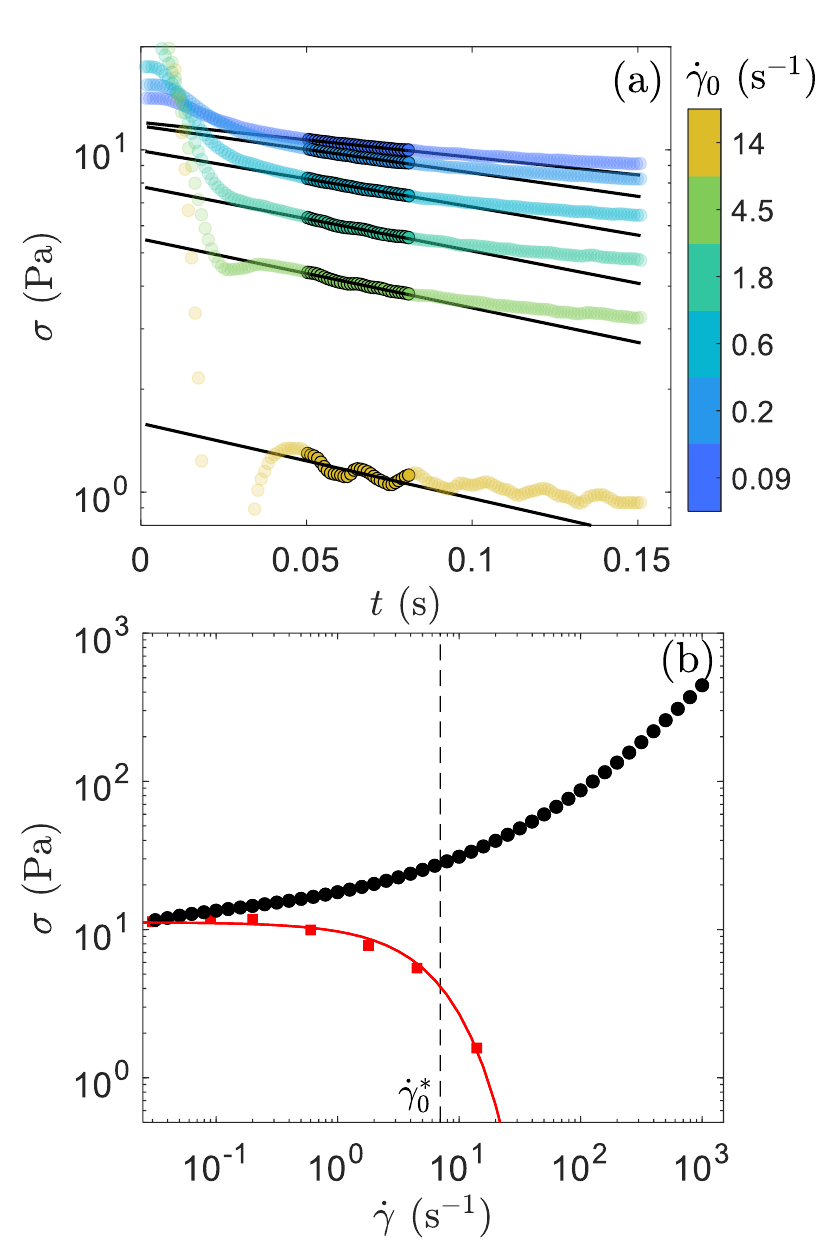}
    \centering
    \caption{Stress jump experiments and the measurements of the elastic stress $\sigma_e$ for $3.2~\%$ carbon black (CB) dispersion. (a) Stress relaxation measured at zero shear rate after applying different final shear rates.  The black lines are the best exponential fits of the stress relaxation data between $5.10^{-2}$ and $8.10^{-2}~\rm s$, which reads: $\sigma(t) = \sigma_e \exp(-t/t_e)$, with $t_e$ and $\sigma_e$ (see Table~\ref{fig:jump} in the Appendix) the slope and intercept in a semi-log plot. (b) Total (black circles) and elastic (red squares) stress measured during the flow curve of $3.2~\%$ CB dispersion (see Fig.~\ref{fig:suppFC}, rate: 1s/point). The red curve is the best fit of the elastic stress with the following function $\sigma_e = \sigma_y \exp(-\dot{\gamma}/\dot{\gamma^*})$, with $\sigma_y = 12~\rm Pa$ and $\dot{\gamma}^* = 7~\rm s^{-1}$.
    }
    \label{fig:jump}
\end{figure}
%%%%%%%%%%%%%%%%%%%%%%%%%%%%%%%%%%%%%%%%%%%%%%%%%%%%%%%%%%%%%%%%%%%%%%%%%%%%%%%%%%%

In that framework, for applied shear rates $\dot{\gamma}_0 < \dot{\gamma}_0^*$, the clusters effectively behave as "sticky" particles and initially organize into a fractal network, as indicated by USAXS ($\alpha = 2$). Following an initial slip phase, likely due to a lubrication layer at the rotor wall, this fractal network progressively densifies, as evidenced by the increase in $\alpha$ and the decrease in conductivity measured by EIS. USV measurements further show that the anti-thixotropic time, $\tau$, marks the end of wall slip, enabling better shear transmission through the gap, which eventually leads to a steady state.

\subsection{Elastic stresses mediate the anti-thixotropic restructuring}

For the initial network formed at $t=0$~s to rearrange and densify, bonds must break, which likely involves straining the network and building up elastic stresses. Thus, the flow of CB suspensions below $\dot{\gamma}_0^*$ should be characterized by the emergence of an elastic contribution to the measured stress. 

To test this hypothesis, we conduct so-called "stress jump" experiments, previously used to separate viscous and elastic stress components in attractive colloidal dispersions~\cite{Dullaert2005,Massaro2020}. In this approach, the stress relaxation is measured immediately after flow cessation. Under shear, the total stress $\sigma$ is assumed to be the sum of hydrodynamic and elastic contributions, i.e., $\sigma = \sigma_h + \sigma_e$. Upon flow cessation, the hydrodynamic stress component $\sigma_h \propto \eta \dot{\gamma}$ vanishes instantly as $\dot{\gamma} \rightarrow 0$, whereas the elastic stress, $\sigma_e$, requires a finite time to relax. Assuming that elastic stress relaxation initially follows an exponential decay, $\sigma_e$ can be determined at $t =0$, i.e., when the flow is stopped. Using this method and a strain-controlled rheometer, we performed a rapid flow sweep on the $3.2~\%$ CB dispersion, stopping the flow at various shear rates. 
Figure~\ref{fig:jump}(a) shows the stress relaxation measured during flow cessation for six different shear rates. The exponential fit of the data allows us to extract $\sigma_e$, which is reported in Figure~\ref{fig:jump}(b) as a function of $\dot{\gamma}$, alongside the total stress measured during the flow sweep. For shear rates $\dot{\gamma} > 20 ~\rm s^{-1}$, no elastic contribution was detected ($\sigma_e =0$), consistent with the flow of isolated clusters governed by hydrodynamic forces~\cite{Bauland2024}. However, for $\dot{\gamma} \leq 20 ~\rm s^{-1}$, the elastic contribution becomes significant, increasing to $\sigma_y$ as the shear rate decreases.

The dependence of $\sigma_e$ with $\dot \gamma$ follows an exponential decay: $\sigma_e = \sigma_y \exp(-\dot{\gamma}/\dot{\gamma^*})$, with $\dot{\gamma}^* = 7~\rm s^{-1}$ [see the red curve in Fig.~\ref{fig:jump}(b)]. This shear rate, characterizing the emergence of elastic stresses during flow, is in excellent agreement with the critical shear rate $\dot \gamma_0^*$ below which anti-thixotropic behavior appears. This finding supports our hypothesis that the restructuring of the fractal network into large agglomerates relies on the elastic deformation of the CB network. 

This interpretation is also consistent with a previous study on CB dispersions, where the orthogonal moduli $G^{\prime}_{\perp}$ and $G^{\prime \prime}_{\perp}$ were measured using orthogonal superposition rheometry during the antithixotropic stress decrease~\cite{Wang2022}. The authors reported a non-monotonic evolution of $G^{\prime}_{\perp}$ and $G^{\prime \prime}_{\perp}$, which initially increased and then decreased, while the shear stress consistently decreased. This difference in timescales between the decrease of these macroscopic variables was attributed to the buildup of structural anisotropy in the vorticity direction. In light of our results, the increase in $G^{\prime}_{\perp}$ and $G^{\prime \prime}_{\perp}$ during antithixotropy could also be interpreted as an increase in the elastic contribution to the stress as the material undergoes shear densification. Furthermore, the peak in $G^{\prime}_{\perp}$ and $G^{\prime \prime}_{\perp}$ roughly corresponds to the inflection point of the shear stress decrease and the anti-thixotropic time, supporting the idea that the material achieves maximum adhesion to the rotor at $\tau$.

\subsection{Scaling of the anti-thixotropic time}

In light of the proposed mechanisms, the exponential increase of the anti-thixotropic time $\tau$ with the shear rate $\dot{\gamma}_0$ can be attributed to the fact that clusters become progressively less "sticky" as the shear rate increases. This leads to a longer time required for the system to form a percolated structure, which diverges as $\dot{\gamma}_0 \rightarrow \dot{\gamma}_0^*$ [Fig.~\ref{fig:rheol}(b)].

This phenomenology is likely a result of the step-flow shear protocol, where the shear rate is applied from high to low values. Indeed, previous studies have shown that the fluidization time for yield stress fluids decreases with increasing shear in creep experiments when starting from rest~\cite{Divoux2010,Divoux2012,Gibaud2010}.

\begin{figure}
    \includegraphics[scale=0.55, clip=true, trim=0mm 0mm 0mm 0mm]{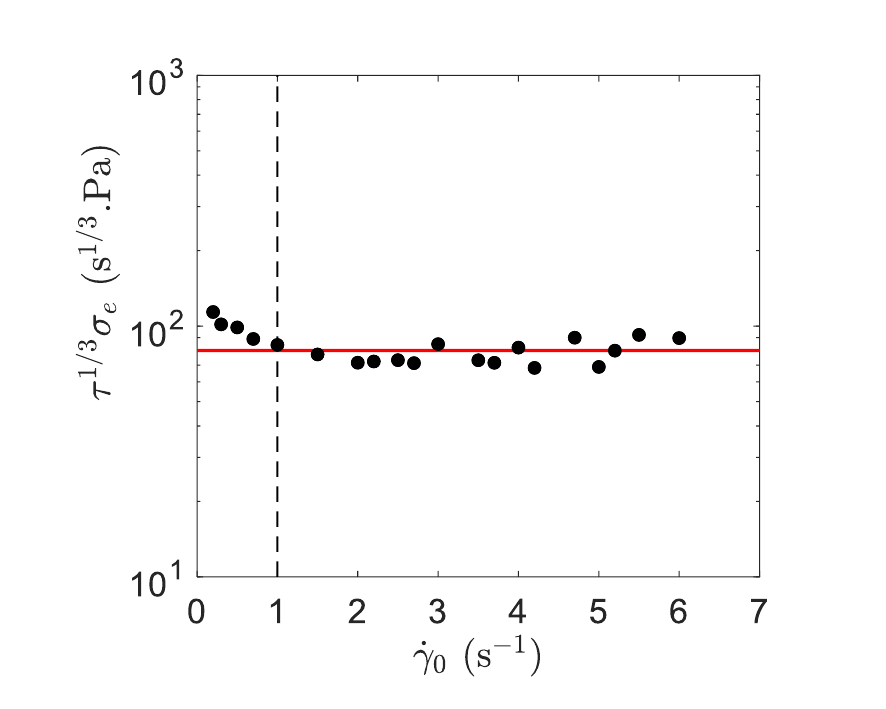}
    \centering
    \caption{Scaling relation between the anti-thixotropic time $\tau$ and the elastic stress $\sigma_e$ as a function of the shear rate. At shear rates below 1~s$^{-1}$, indicated by the vertical dash line, slip at the rotor wall remains significant. 
    }
    \label{fig:scaletau}
\end{figure}

To further rationalize this dynamics, we compare the elastic stress $\sigma_e$ (shown in Fig.~\ref{fig:jump}), which decreases as $\dot{\gamma}^*_0$ is approached, with the anti-thixotropic time $\tau$ (shown in Fig.~\ref{fig:rheol}), which diverges near $\dot{\gamma}^*_0$. As shown in Fig.~\ref{fig:scaletau}, $\sigma_e$ and $\tau$ are correlated and follow the scaling relation $\tau^{1/3} = A/\sigma_e$, where $A = 80~\mathrm{Pa~s^{1/3}}$ is a constant. This scaling demonstrates that the amplitude of the elastic stress drives the dynamics of the anti-thixotropic restructuring. Given the difference in exponents between $\sigma_e$ and $\tau$, the value of $\tau$ is extremely sensitive to $\sigma_e$. This scaling also suggests the potential to define a dimensionless number for anti-thixotropy analogous to the Mnemosyne Number $My$ used for thixotropy. However, evaluating such a number would require a broader dataset across different systems and concentrations, which is beyond the scope of this study.

%%%%%%%%%%%%%%%%%%%%%%%%%%%%%%%%%%%%%%%%%%%%%%%%%%%%%%%%%%%%%%%%%%%%%%%%%%%%%%%%%%
%%%%%%%%%%%%%%%%%%%%%%%%%%%%%%%%%%%%%%%%%%%%%%%%%%%%%%%%%%%%%%%%%%%%%%%%%%%%%%%%%%%

%%%%%%%%%%%%%%%%%%%%%%%%%%%%%%%%%%%%%%%%%%%%%%%%%%%%%%%%%%%%%%%%%%%%%%%%%%%%%%%
\begin{figure}[!t]
    \includegraphics[scale=0.55, clip=true, trim=0mm 0mm 0mm 0mm]{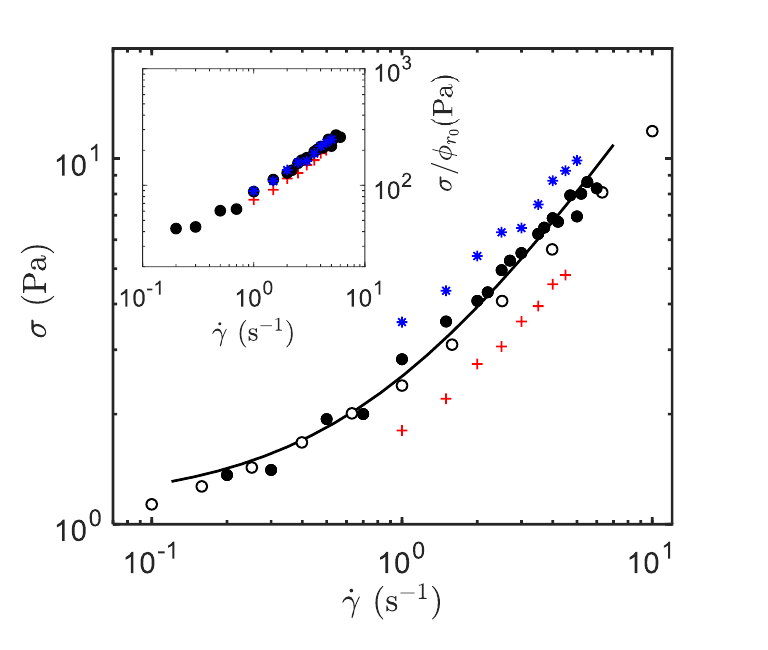}
    \centering
    \caption{Flow curves of carbon black (CB) dispersions with various volume fractions in the anti-thixotropic regime. Full markers are the final stress $\sigma_{\infty}$ taken at $t=10^4~\rm s$ after imposing a flow step-down, plotted as a function of the applied shear rate $\dot{\gamma}_0$ (denoted $\dot{\gamma}$ in x-axis for generality). Marker style and color codes for the volume fraction of CB $\phi_{r_0}=$ 2.4 ($+$), 3.2 ($\bullet$), 4.1 ($\ast$). The black curve is the best fit of the $3.2~\rm \%$ data with the Bingham model, yielding $\sigma_y = 1.2~\rm Pa$ and $\eta_{bg} = 1.4~\rm Pa.s$. Empty black markers are the flow curve of the 3.2\% CB dispersion measured by flow sweep, after shearing the dispersion at $\dot{\gamma}_0 = 1~\rm s^{-1}$ during $10^4~\rm s$ (see text for further details). Inset: stress normalized by the CB volume fraction versus the shear rate for $\phi_{r_0}=$ 2.4 ($+$), 3.2 ($\bullet$), 4.1 ($\ast$).
    }
    \label{fig:fc}
\end{figure}
%%%%%%%%%%%%%%%%%%%%%%%%%%%%%%%%%%%%%%%%%%%%%%%%%%%%%%%%%%%%%%%%%%%%%%%%%%%%%%%

\subsection{Lack of memory upon reaching a steady state}

After discussing the mechanisms behind the long-lasting dynamics of the anti-thixotropic stress response, we now focus on the steady state. To define this state, we arbitrarily use the stress value reached at $t = 10^4~\rm s$, which corresponds to $t \gg \tau$ [Fig.~\ref{fig:rheol}(a)], and denote it $\sigma_{\infty}$. This stress value is reported in Figure~\ref{fig:fc} against the applied shear rate $\dot{\gamma}_0$ (denoted $\dot{\gamma}$) to construct an effective flow curve. The effective flow curve is well described by a Bingham model, which is expressed as $\sigma = \sigma_y + \eta_{bg}\dot{\gamma}$, where the yield stress $\sigma_y = 1.2~\rm Pa$ is much lower than the value $\sigma_y = 14.7~\rm Pa$ measured from a rapid flow sweep (see Fig.~\ref{fig:suppFC}). This suggests that the yield stress of the 3.2~\% CB dispersion can be tuned between $1.2$ and $14.7~\rm Pa$, depending on the shear history. Specifically, if the flow step-down is interrupted by bringing the dispersion at rest ($\dot{\gamma}=0$) at $t\ll\tau$, then $\sigma_y \simeq 14.7$~Pa; if $t\gg\tau$, then $\sigma_y \simeq 1.2$~Pa.
Moreover, the Bingham model implies that the steady-state flow behavior of CB dispersions is characterized by a single viscosity, $\eta_{bg}$, across the range $\dot{\gamma} \in [0.1-10]~\mathrm{s^{-1}}$. This indicates that the structure formed after the anti-thixotropic stress response —composed of dense, large agglomerates— remains stable over this range of shear rates. Indeed, if the agglomerate size varied along the flow curve, we would expect the viscosity $\eta_{bg}$ to depend on the shear rate, $\eta_{bg} = \eta_{bg}(\dot{\gamma}$)~\cite{wang2019}.

\begin{figure}[t!]
    \includegraphics[scale=0.45, clip=true, trim=15mm 0mm 5mm 0mm]{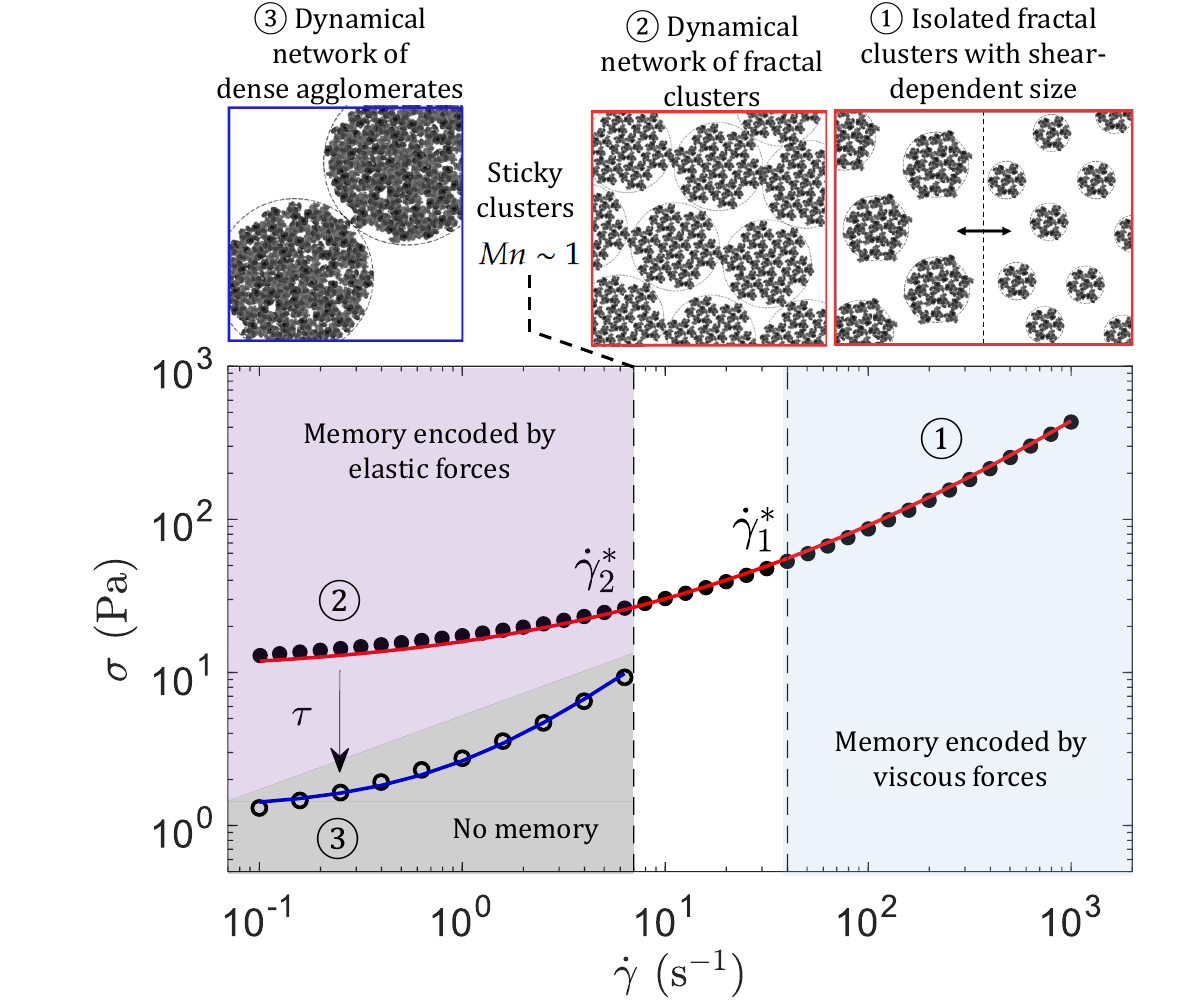}
    \centering
    \caption{Graphical summary of the flow properties of CB dispersions. Data corresponds to the dispersion with $\phi_{r_0} = 3.2~\%$: full black markers correspond to a flow sweep ($\Delta t = 1~\rm s$ per point) while empty markers correspond to the final stress values reached at $t = 10^4~\rm s$ following a step flow of shear rate for $\dot{\gamma}_0 < 7~\rm s^{-1}$. Red curve is the best fit with the three components model~\cite{Caggioni2020}, which reads $\sigma = \sigma_y + \sigma_y (\dot{\gamma}/\dot{\gamma}^*)^{(1/2)} + \eta_{r_0}\dot{\gamma}$, with $\sigma_y = 14.7~\rm Pa$, $\dot{\gamma}^* = 6.7~\rm s^{-1}$, $\eta_{r_0} = 0.25~\rm Pa.s$. Blue curve is the best fit with the  Bingham model, which reads $\sigma = \sigma_y + \eta_{bg}\dot{\gamma}$, with $\sigma_y = 1.2~\rm Pa$ and $\eta_{bg} = 1.4~\rm Pa.s$.  \textcircled{1} The equilibrium size of fractal clusters is determined by the stress of the background fluid. \textcircled{2} The maximum clusters size is determined by their dynamic percolation at $\dot{\gamma}_1^*$ corresponding effective volume fraction in clusters $\phi_{eff} \simeq 0.4$. During a fast flow sweep, the yield stress value ($\sigma_y = 14.7~\rm Pa)$ is inherited from the percolation point. \textcircled{3} Below a second critical shear rate $\dot{\gamma}_2^*$ (denoted $\dot{\gamma}_0^*$ in the body text), where clusters become effectively sticky, dispersion exhibit anti-thixotropy corresponding to the transition from a network of fractal clusters to dense agglomerate within a characteristic time $\tau$.
    }
    \label{fig:bilan}
\end{figure}

To test this hypothesis, we conducted a flow step-down at $\dot{\gamma}_0 = 1~\rm s^{-1}$ for $10^4~\rm s$, as previously described, followed by a flow sweep between $0.1 \leq \dot{\gamma} \leq 10~\rm s^{-1}$. The result is shown in Figure~\ref{fig:fc}, where the effective flow curve, built from the final stress values of the flow steps (black markers), and the actual flow curve (orange markers) overlap. This indicates that the anti-thixotropic behavior leads to a single structuring of the dispersions. This result is further supported by the final scattering curves at $t = 10^3~\rm s$, which are identical for $\dot{\gamma}_0 = 0.5$ and $1~\rm s^{-1}$, with $\alpha = 3.2$ [Fig.~\ref{fig:SAXS}(d)]. Since the agglomerates are no longer fractal, we expect the viscosity of the dispersion to scale linearly with the volume fraction of primary CB particles $\phi_{r_0}$. This is confirmed in the inset of Figure~\ref{fig:fc}, where the effective flow curves can be rescaled by normalizing the stress $\sigma_{\infty}$ by $\phi_{r_0}$. These results support our hypothesis that the anti-thixotropic stress decay leads to a single structure, likely defined by a single agglomerate size. Consequently, at low shear rates, for $\dot{\gamma} < \dot{\gamma}_0^*$, the shear flow memory stems from the transient dynamics of the anti-thixotropic behavior. 

Our results are reminiscent of the "flow-switched bistability" reported for attractive colloidal dispersions with non-interacting filler particles~\cite{Jiang2022,Larsen2024}. In those studies, after a step-down in stress, it was observed that the presence of non-Brownian particles embedded in a gel matrix composed of attractive colloidal particles allowed the colloids to segregate into agglomerates under moderate shear stresses. This segregation led to the 'liquefaction' of the dispersion, characterized by a vanishingly small yield stress. Interestingly, the agglomerate or "blob" size was found to be independent of the applied stress, similar to our results when controlling the shear rate. This suggests a comparable mechanism in both cases. In Figure~\ref{fig:rheol}(b), we have shown that the anti-thixotropic time increases as the volume fraction of primary particles decreases. Based on this, it can be hypothesized that the effect of filler particles reported in~\cite{Jiang2022,Larsen2024} is an excluded volume effect, concentrating the attractive colloidal particles. In other words, the addition of large non-interacting particles likely concentrates the attractive Brownian particles, bringing the anti-thixotropic time into measurable timescales. Consequently, anti-thixotropy may be an intrinsic property of sheared colloidal gels, with its characteristic timescale depending on the volume fraction of particles and, by extension, on the filler content.

%%%%%%%%%%%%%%%%%%%%%%%%%%%%%%%%%%%%%%%%%%%%%%%%%%%%%%%%%%%%%%%
\section{Conclusion}

In this study, we investigated the anti-thixotropic response of CB dispersions to flow step-down tests. Our results reveal that anti-thixotropy occurs below a critical shear rate of about $\dot{\gamma}^*_0 \approx 7~\mathrm{s}^{-1}$. In this regime, following a flow step-down, the stress decays over a characteristic time $\tau$ and then reaches a steady state. Notably, $\tau$ increases exponentially with the applied shear rate.

Using a combination of Ultrasonic Speckle Velocimetry, Ultra Small Angle X-ray Scattering, and Electrical Impedance Spectroscopy measurements, we have shown that this behavior is driven by a combination of wall slip at short times, followed by slow structural rearrangements within the dispersion. Immediately after the flow step-down, fractal clusters of CB particles, formed during preshear, percolate into a dynamical fractal network. This network then undergoes further restructuring under shear, ultimately forming a dynamical network of weakly connected large and dense aggregates that are no longer fractal. 
In the anti-thixotropic regime, the driving force for this transition is the presence of elastic stress, such that the time $\tau$ required to reach the stationary anti-thixotropic state is inversely proportional to the elastic stress. More precisely, $\tau^{1/3} = {A}/{\sigma_e}$. 
Notably, the steady state reached after prolonged shear does not depend on the applied shear rate, suggesting that the dispersion retains no memory of the flow history. The critical shear rate $\dot{\gamma}^*_0 \approx 7~\mathrm{s}^{-1}$, corresponding to a Mason number $Mn\simeq 1$,  marks the threshold where clusters become effectively sticky and the emergence of an elastic contribution to the shear stress, enabling the deformation and the subsequent compaction of the microstructure. Our multi-method approach has provided a detailed microstructural scenario for the anti-thixotropic behavior in CB dispersions, paving the way for a deeper understanding of the structuring of colloidal dispersions under moderate shear.

Finally, we can provide a comprehensive overview of the shear-induced structuring of CB dispersions by combining the present work with our previous study~\cite{Bauland2024}. Figure~\ref{fig:bilan} shows the flow curve of a 3.2\% CB dispersion alongside a sketch of the microstructure evolution over four decades in shear rates. In sketch \textcircled{1}, the high shear rate region of the flow curve is characterized by a hydrodynamic regime, where the dispersion consists of isolated clusters. The equilibrium size of these clusters is directly influenced by the viscous forces~\cite{Bauland2024}. In this regime, steady-state is achieved rapidly (within a few seconds), and the flow memory is encoded by the hydrodynamic stresses $\sigma_h$ of the background fluid. The characteristic shear rate $\dot{\gamma}_1^*$ marks the lower bound of this hydrodynamic regime, corresponding to the 'dynamic percolation' of fractal clusters, which determines their maximum size, as depicted in sketch \textcircled{2}.  
In the limit $\dot{\gamma} \rightarrow 0$, this fractal structure is conserved as long as the shear time or total strain remains low. However, below a second characteristic shear rate $\dot{\gamma}_2^*$ (denoted $\dot{\gamma}_0^*$ in the body text), the clusters become effectively sticky, and time-dependent effects, namely anti-thixotropy, emerge. Over time, the dispersion transitions from a dynamic network of fractal clusters —whose structure is inherited from $\dot{\gamma}_1^*$— to large, dense agglomerates, as depicted in sketch \textcircled{3}. In this regime, flow memory is encoded by elastic stresses $\sigma_e$ that deform and densify the fractal network \textcircled{2}, leading to a continuous decrease in viscosity and yield stress.  Once a steady state is reached, the system no longer retains memory of the previous flow history, as a single structure is formed that is independent of the shear rates for $\dot{\gamma} < \dot{\gamma}_2^*$.

As a perspective, the structuring of attractive colloidal dispersions, controlled by flow rate and duration (Fig.~\ref{fig:bilan}), presents opportunities for exploitation. Rapidly quenching the flowing dispersion into a solid state through flow cessation could allow the formation of gels with varied structures and mechanical properties, without changing the formulation. These experiments could reveal how flow memory is imprinted in gels and find direct applications in injectable hydrogel or 3D printing processes.

%%%%%%%%%%%%%%%%%%%%%%%%%%%%%%%%%%%%%%%%%%%%%%%%%%%%%%%%%%%%%%%%%%%%%
\section*{Author Contributions}
JB, AP, and TG conducted the rheo-USAXS experiments. JB and SM performed the USV experiments. JB, TD, and GL carried out the Rheo-EIS experiments. Each author analyzed the data from their respective experiments. All authors contributed to data discussions and manuscript editing. JB and TG wrote the article. TG designed and supervised the project.
 
\section*{Conflicts of interest}
There are no conflicts to declare.

\section*{Acknowledgements}
The authors are especially grateful to the ESRF for beamtime at the beamline ID02 (proposal SC-5236) and Theyencheri Narayanan for the discussions and technical support during the USAXS measurements. 
This work was supported by the Région Auvergne-Rhône-Alpes ``Pack Ambition Recherche", the LABEX iMUST (ANR-10-LABX-0064) of Université de Lyon, within the program "Investissements d'Avenir" (ANR-11-IDEX-0007), the ANR grants (ANR-18-CE06-0013 and ANR-21-CE06-0020-01) and European Union’s Horizon Europe Framework Programme (HORIZON) under the Marie Skłodowska-Curie Grant Agreement 101120301. This work benefited from meetings within the French working group GDR CNRS 2019 ``Solliciter LA Matière Molle" (SLAMM).

%\bibliography{library}

%\clearpage 

\section{Appendix}
\label{sec:app}
\subsection{Rheological measurements}

\begin{figure}[t!]
    \includegraphics[scale=0.5, clip=true, trim=0mm 0mm 0mm 0mm]{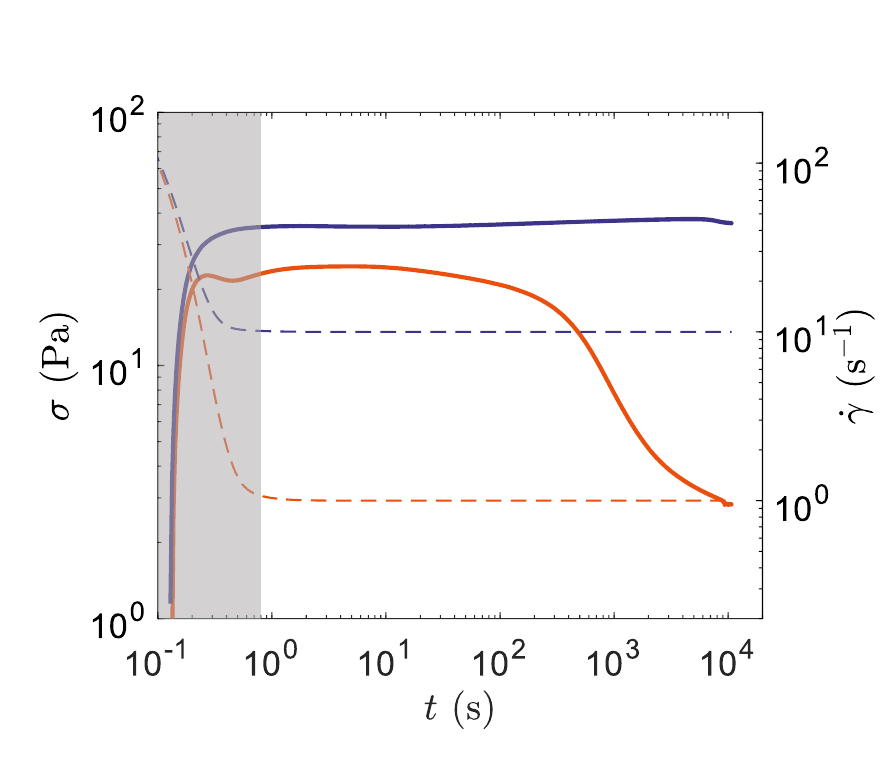}
    \centering
    \caption{Temporal evolution of the stress $\sigma$ (solid lines) and shear rate $\dot{\gamma}$ (dotted lines) following a flow step-down from $\dot{\gamma}=500~\rm s^{-1}$ to $\dot{\gamma}_0 = 10$ (blue) and $1~\rm s^{-1}$ (orange). Data corresponds to a CB dispersion with a volume fractions $\phi_{r_0} = 3.2~\%$.}
    \label{fig:supFSD2}
\end{figure}

\begin{figure*}[t!]
    \includegraphics[scale=0.58, clip=true, trim=0mm 0mm 0mm 0mm]{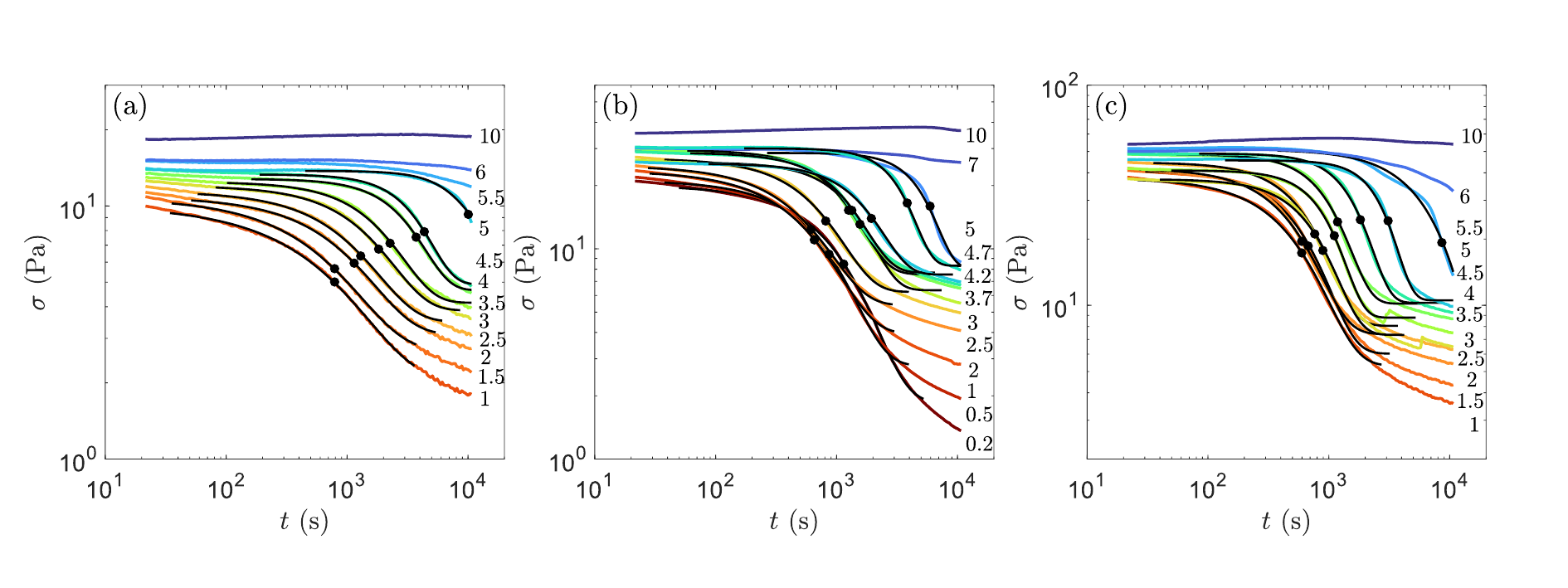}
    \centering
    \caption{Temporal evolution of the stress response $\sigma$ following a flow step-down from $\dot{\gamma}=500~\rm s^{-1}$ to $\dot{\gamma}_0$, for carbon black dispersions with volume fractions $\phi_{r_0}$ of (a) 2.4, (b) 3.2 and (c) 4.1~\%. The values of $\dot{\gamma}_0$ are displayed on the right side of each panel. Black lines are the best exponential fits $\sigma(t) = \sigma_0 + \sigma_1.exp[-(t / \tau)^{\beta}]$, with $t = \tau$ indicated as a black dot.}
    \label{fig:supFSD}
\end{figure*}

During flow step-down experiments, the stress-controlled rheometers successfully imposed $\dot{\gamma}_0$ in about $0.8~\rm s$. Consequently, data for $t \leq 0.8~\rm s$ are not considered, as evidenced in Figure~\ref{fig:supFSD2}. 
Additionally, flow step-down experiments were investigated at three volume fractions of CB particles, $\phi_{r_0} =$ 2.4, 3.2 and 4.1~\%. Figure~\ref{fig:supFSD}(a)-(c) displays the complete data set, corresponding to Figure~\ref{fig:rheol}(b)-(c) in the main text.

%%%%%%%%%%%%%%%%%%%%%%%%%%%%%%%%%%%%%%%%%%%%%%%%%%%%%%%%%%%%%%%%%%%%%%%%%%%%%%%%%%%%
%%%%%%%%%%%%%%%%%%%%%%%%%%%%%%%%%%%%%%%%%%%%%%%%%%%%%%%%%%%%%%%%%%%%%%%%%%%%%%%%%%%%

%%%%%%%%%%%%%%%%%%%%%%%%%%%%%%%%%%%%%%%%%%%%%%%%%%%%%%%%%%%%%%%%%%%%%%%%%%%%%%%%%%%%
%%%%%%%%%%%%%%%%%%%%%%%%%%%%%%%%%%%%%%%%%%%%%%%%%%%%%%%%%%%%%%%%%%%%%%%%%%%%%%%%%%%%

\subsection{Ultrasonic speckle velocimetry}

\begin{figure}[t!]
    \includegraphics[scale=0.47, clip=true, trim=0mm 0mm 0mm 0mm]{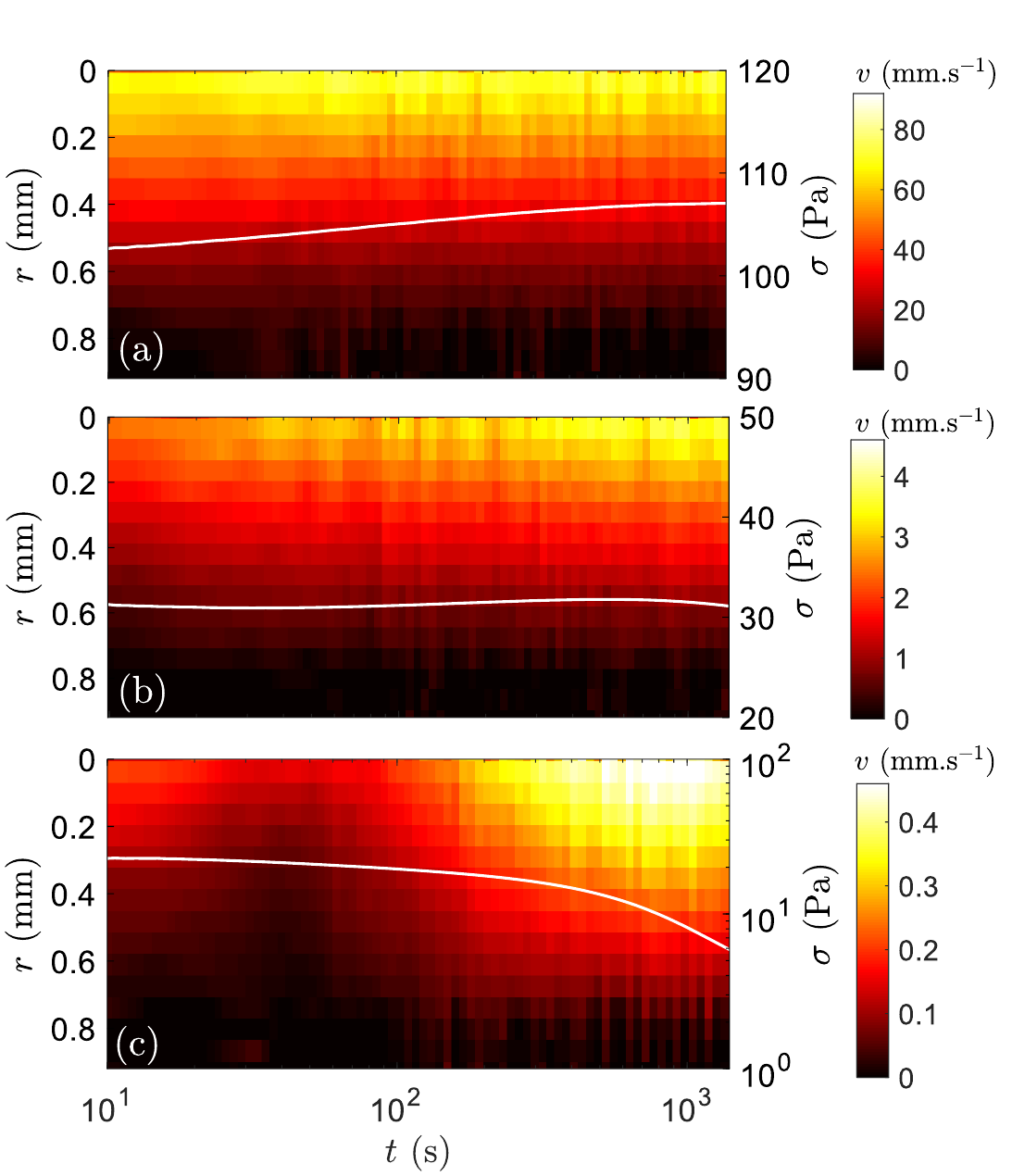}
    \centering
    \caption{Spatiotemporal diagrams of the velocity data $v(r,t)$ as a function of position $r$ and time $t$ for (a) $\dot{\gamma}_0 = 100~\rm s^{-1}$, (b) $\dot{\gamma}_0 = 5~\rm s^{-1}$ and (c) $\dot{\gamma}_0 =  0.5~\rm s^{-1}$. The stress response $\sigma(t)$ is displayed on the right vertical axis.}
    \label{fig:supUSV2}
\end{figure}

Figure~\ref{fig:supUSV2} displays the spatiotemporal diagrams of the velocity data for $\dot{\gamma}_0 = 100$, $5$ and $0.5~\rm s^{-1}$.

%%%%%%%%%%%%%%%%%%%%%%%%%%%%%%%%%%%%%%%%%%%%%%%%%%%%%%%%%%%%%%%%%%%%%%%%%%%%%%%%%%%%

\subsection{Small angle X-ray scattering}

\begin{figure}[t!]
    \includegraphics[scale=0.50, clip=true, trim=0mm 0mm 0mm 0mm]{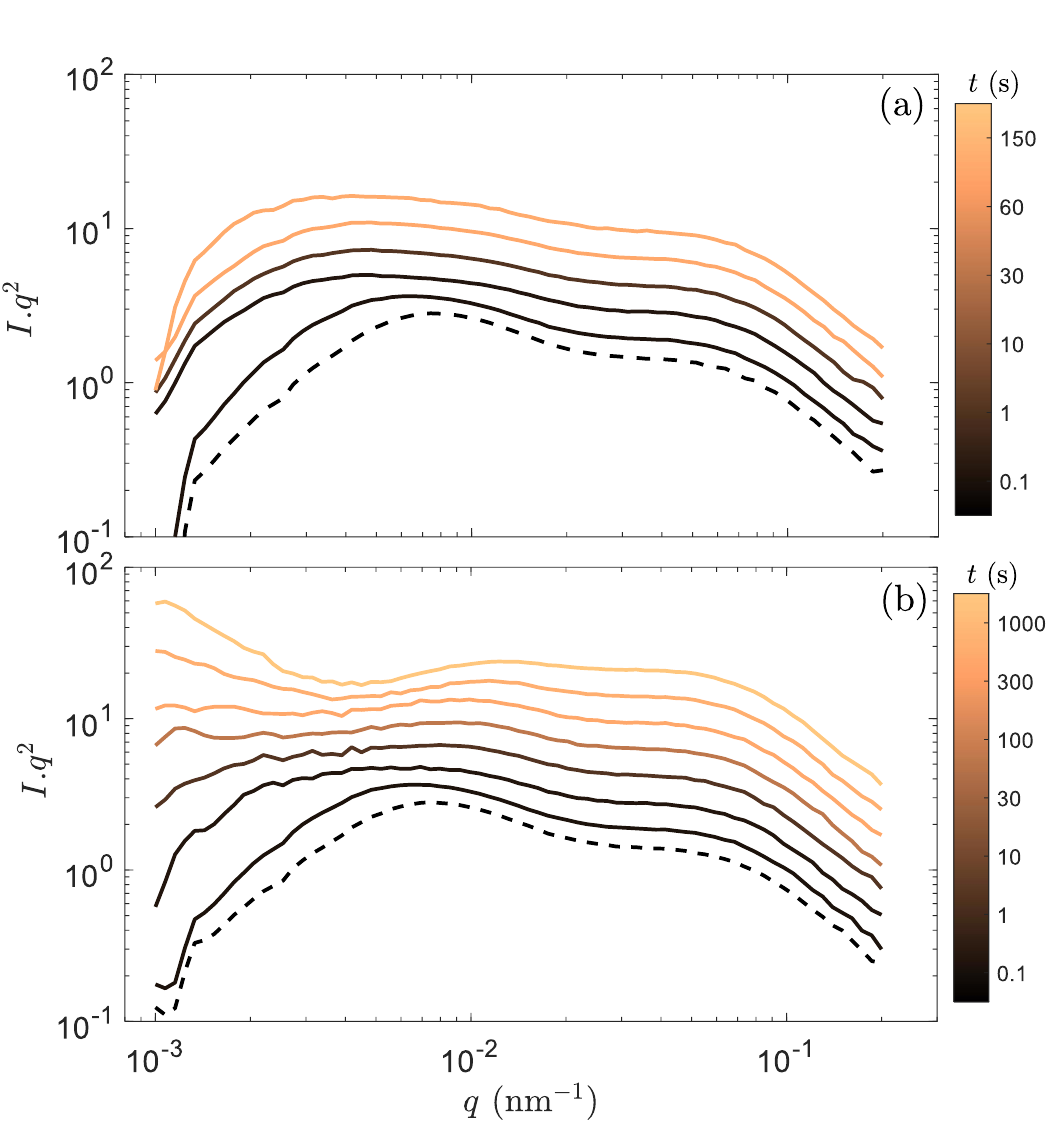}
    \centering
    \caption{Rheo-USAXS of the 3.2 \% of carbon black (CB) dispersion during flow step-down. (a)-(b) Kratky plots $I.q^2$ vs.~$q$ of the scattering profiles acquired at different times after imposing a shear rate of (a) $\dot{\gamma}_0 = 100~\rm s^{-1}$ and (b) and $0.5~\rm s^{-1}$. Black dotted curves are the scattering profile of the dispersion during the preshear step at $\dot{\gamma} = 500~\rm s^{-1}$.}
    \label{fig:supSAXS}
\end{figure}

Figures~\ref{fig:supSAXS}(a)-(b) display the evolution of the 1D scattering intensity $I(q)$ for the 3.2~\% CB dispersion under application of $\dot{\gamma}_0 = 100$ and $0.5~\rm s^{-1}$. 

To provide a quantitative analysis of the evolution of the low-$q$ part of the spectrum, the intensity $I(q)$ in the range $10^{-3} \leq q \leq 10^{-2}~\rm nm^{-1}$ was adjusted by either a Guinier-Porod~\cite{Hammouda2010} (2) or a simple power-law model (3) depending on the shape of the data:   

\begin{equation}
\begin{array}{l}
\begin{array}{ll}
P(q) & = \left\{
\begin{array}{lll}
\frac{G}{q^s}exp\bigg(\frac{-q^2Rg^2}{3-s}\bigg) \qquad \ \ q \leq q_1, \\
\frac{D}{q^m} \qquad \qquad \qquad \qquad  q \geq q_1 \\
\end{array}
\right.
\end{array}
\end{array}
\end{equation}

\begin{equation}
P(q) = \frac{K}{q^{\alpha}} \\
\end{equation}

Indeed, performing flow step-down leads to an increase of the cluster size that, for the lowest shear rates, exceeds the upper detection range of the USAXS set-up. Accordingly, the low-$q$ structuring of the 3.2~\% CB dispersion is sequentially captured by the Guinier-Porod and the power-law model (see Fig.~\ref{fig:SAXS} in the main text).

\begin{figure}[t!]
    \includegraphics[scale=0.45, clip=true, trim=0mm 0mm 0mm 0mm]{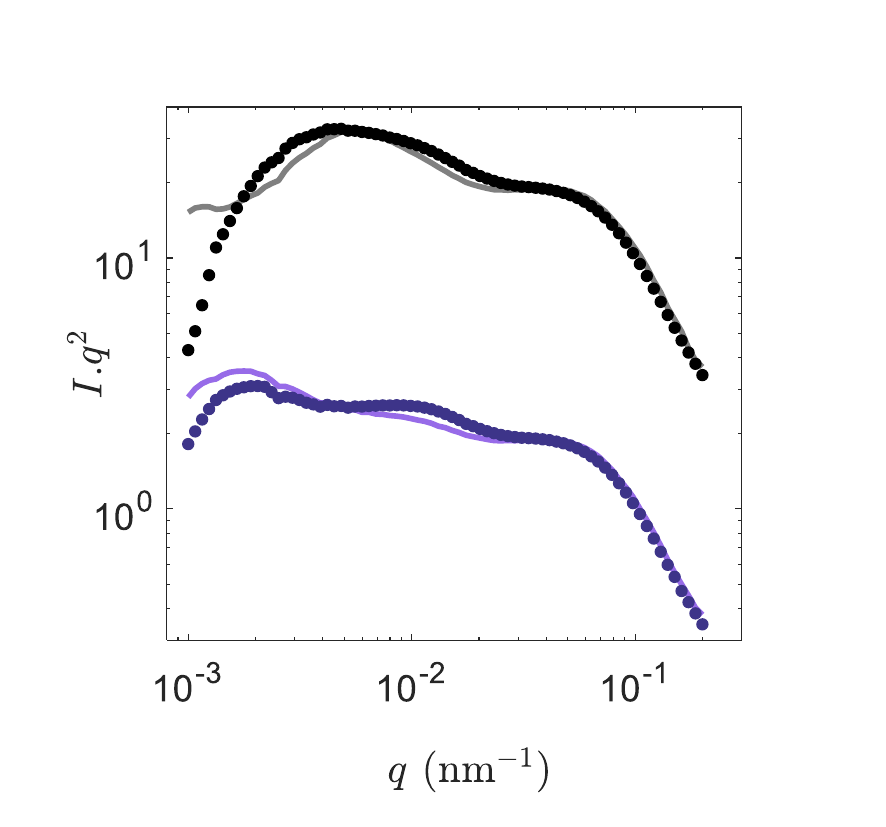}
    \centering
    \caption{Comparison of the scattering curves at $\dot{\gamma}=10$ and $100~\rm s^{-1}$ after a flow step-down from $\dot{\gamma} = 500~\rm s^{-1}$ to $\dot{\gamma}_0$ (markers) or a flow sweep with $\Delta t = 1~\rm s$ (solid lines).}
    \label{fig:supSAXS2}
\end{figure}

In Figure~\ref{fig:supSAXS2}, we compare the scattering intensity of the 3.2~\% dispersion measured at $\dot{\gamma}=10$ and $100~\rm s^{-1}$ under two different protocols: (i) a flow step-down (markers) and (ii) a flow sweep (solid lines). For both shear rates, data overlap, indicating that for $\dot{\gamma} > \dot{\gamma}_0^*$, the same steady state is achieved, regardless of the flow history.

%%%%%%%%%%%%%%%%%%%%%%%%%%%%%%%%%%%%%%%%%%%%%%%%%%%%%%%%%%%%%%%%%%%%%%%%%%%%%%%%%%%%%%
%%%%%%%%%%%%%%%%%%%%%%%%%%%%%%%%%%%%%%%%%%%%%%%%%%%%%%%%%%%%%%%%%%%%%%%%%%%%%%%%%%%%%%

%%%%%%%%%%%%%%%%%%%%%%%%%%%%%%%%%%%%%%%%%%%%%%%%%%%%%%%%%%%%%%%%%%%%%%%%%%%%%%%%%%%%%%%%%%%%%%
%%%%%%%%%%%%%%%%%%%%%%%%%%%%%%%%%%%%%%%%%%%%%%%%%%%%%%%%%%%%%%%%%%%%%%%%%%%%%%%%%%%%%%%%%%%%%

\subsection{Modeling of complex impedance}

\begin{figure}[t!]
    \includegraphics[scale=0.5, clip=true, trim=0mm 0mm 0mm 0mm]{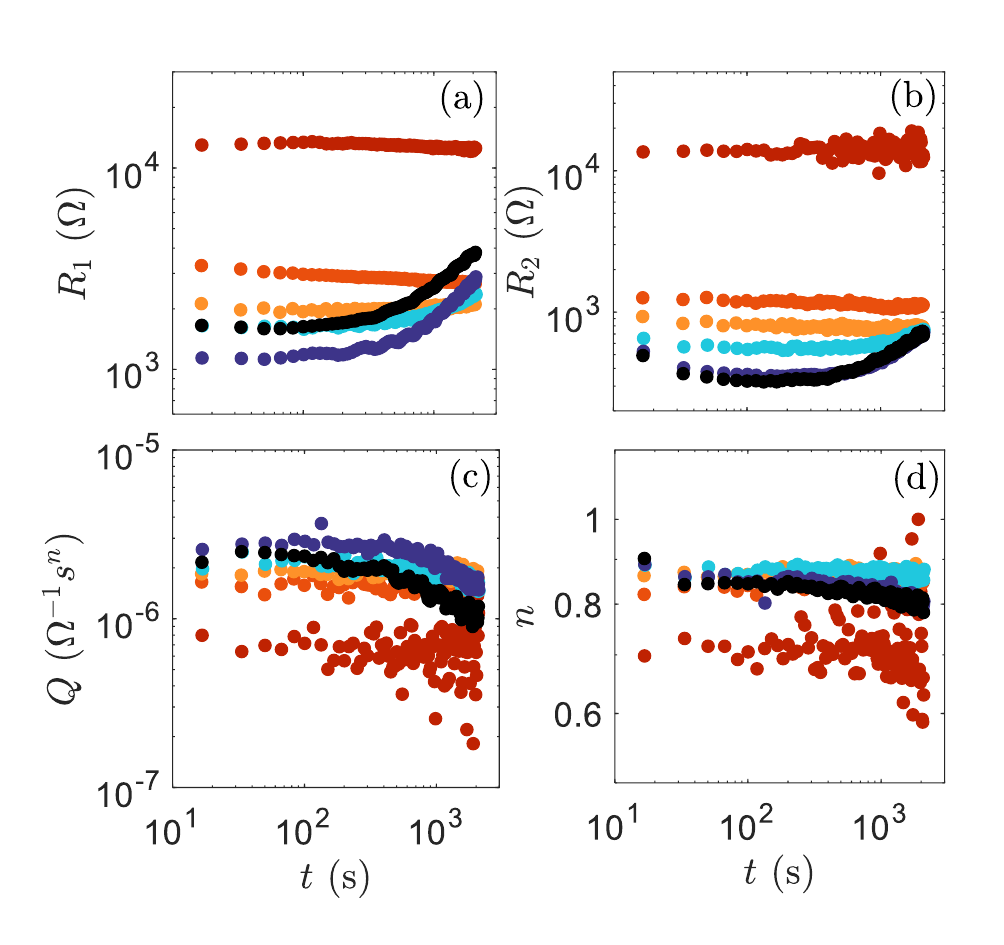}
    \centering
    \caption{Modeling of the complex impedance of the 3.2 \% of carbon black (CB) dispersion using the electrical model displayed in inset of Figure~\ref{fig:elec}(b), composed of a resistance $R_1$ in parallel with a second resistance $R_2$ and a constant phase element of parameters $Q$ and $n$. Colors code for the applied shear rate $\dot{\gamma}_0 = 100$ (black), 10 (blue), 5 (cyan), 2 (light orange), 1 (dark orange) and $0.5~\rm s^{-1}$ (red). Panels (a)-(d) display the temporal evolution of the model parameters during shear.
    }
    \label{fig:supconducti}
\end{figure}

As determined from EIS measurements, the complex impedance $Z^*(f_e)$ is a function of the frequency $f_e$ of the applied oscillatory tension. It can be captured by the following electrical model~\cite{Legrand2022} [see scheme in Fig.~\ref{fig:elec}(b) of the main text]: 

\begin{equation}
    Z^*(f_e) = R_2 \frac{1 + R_2Q(i2\pi f_e)^n}{1 + (R_1 + R_2)Q(i2\pi f_e)^n}
\end{equation}

\noindent with a resistance $R_1$ in parallel with a second resistance $R_2$ and a constant phase element of parameters $Q$ and $n$~\cite{Cole1928}. The constant phase element generalizes the behavior between a perfect capacitor ($n=1$) and a resistance ($n=0$). In this model, $R_1$ is associated with the resistance of CB particles, while $R_2$ and the constant phase element account for the ionic resistance of the solvent and the charge accumulation at the electrode surfaces, respectively~\cite{Legrand2022}. Figure~\ref{fig:supconducti}(a)-(c) displays the models parameters as a function of time and for different applied shear rates.

The resistance $R_1$ associated with the resistance of CB particles was converted into an electrical conductivity $\Sigma_1$ according $\Sigma_1 = k / R_1$, with $k = 0.013~\pm 0.005$~cm$^{-1}$ the cell constant of the EIS set-up~\cite{Legrand2022}.  

%%%%%%%%%%%%%%%%%%%%%%%%%%%%%%%%%%%%%%%%%%%%%%%%%%%%%%%%%%%%%%%%%%%%%%%%%%%%%%   
%%%%%%%%%%%%%%%%%%%%%%%%%%%%%%%%%%%%%%%%%%%%%%%%%%%%%%%%%%%%%%%%%%%%%%%%%%%%%% 

\subsection{Flow cessation experiments}

\begin{figure}[t!]
    \includegraphics[scale=0.5, clip=true, trim=0mm 0mm 0mm 0mm]{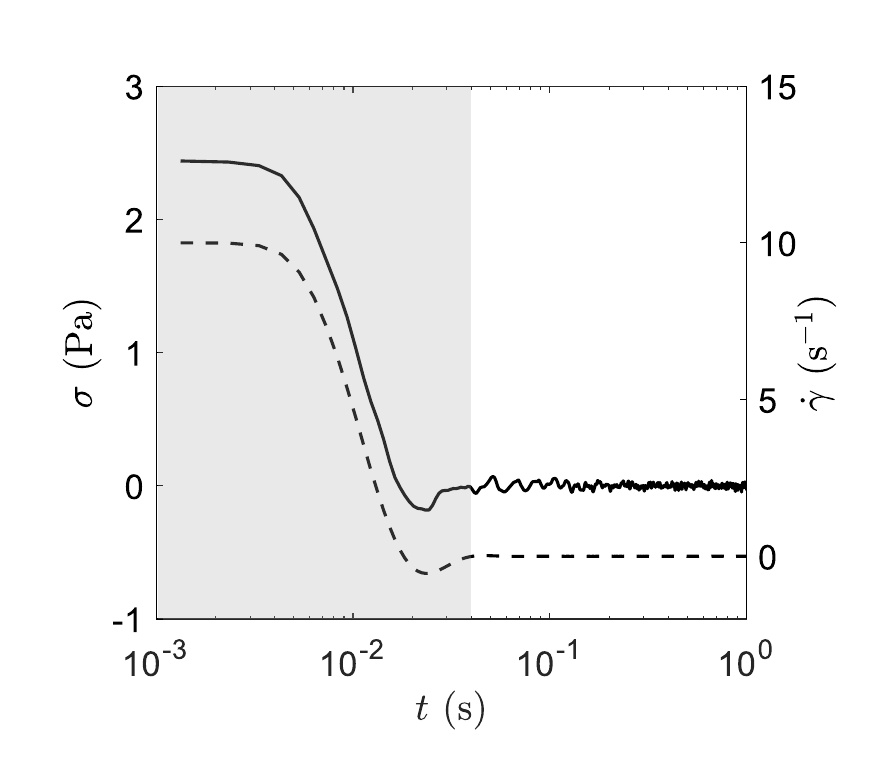}
    \centering
    \caption{Shear stress (solid line) and shear rate (dotted line) vs. time during flow cessation experiments performed on the background fluid (mineral oil) with a strained-controlled rheometer. From $\dot{\gamma} = 10~\rm s^{-1}$, the zero shear rate condition was imposed in about $40~\rm$ ms.}
    \label{fig:suppjump}
\end{figure}

To determine the elastic contribution to the shear stress, we perform flow cessation experiments or so-called "stress jump", and measured the relaxation of the stress as depicted in Figure~\ref{fig:jump}(a) in the main text. We use a strain-controlled rheometer (ARES-G2, TA instruments), allowing to impose rapidly $\dot{\gamma} = 0$~s$^{-1}$ in $40$~ms, as depicted in Figure~\ref{fig:suppjump}. Accordingly, the stress relaxation data were fitted with an exponential model in the range $5.10^{-2} \leq t \leq 8.10^{-2}~\rm s$. The exponential model reads $\sigma(t) = \sigma_e \exp(-t/T)$, with $T$ and $\sigma_e$ the slope and intercept in a semi-log plot. The fit parameters are displayed in Table~\ref{table1} for different shear rates. 

%%%%%%%%%%%%%%%%%%%%%%%%%%%%%%%%%%%%%%%%%%%%%%%%%%%%%%%%%%%%%%%%%%%%%%%%%%%%%%   
%%%%%%%%%%%%%%%%%%%%%%%%%%%%%%%%%%%%%%%%%%%%%%%%%%%%%%%%%%%%%%%%%%%%%%%%%%%%%% 

%%%%%%%%%%%%%%%%%%%%%%%%%%%%%%%%%%%%%%%%%%%%%%%%%%%%%%%%%%%%%%%%%%%%%%%%%%%%%% 
\begin{table}[h!]
        \begin{tabular}{ c c c }
            \hline
           $\dot{\gamma}$ (s$^{-1}$) & $\sigma_e$ (Pa) & $t_e$ (s) \\
             \hline
                14 & 1.6 & 0.20 \\
                4.5 & 5.5 & 0.22 \\
                1.8 & 7.8 & 0.23 \\
                0.6 & 9.9 & 0.26 \\
                0.2 & 11.7 & 0.31 \\
                0.09 & 12 & 0.42 \\
                0.03 & 11.3 & 0.60 \\
                0.01 & 9.5& 1 \\
             \hline
        \end{tabular}
    \caption{Parameters corresponding to the exponential fit of the stress relaxation data in Figure~\ref{fig:jump}}
    \label{table1}
\end{table}
%%%%%%%%%%%%%%%%%%%%%%%%%%%%%%%%%%%%%%%%%%%%%%%%%%%%%%%%%%%%%%%%%%%%%%%%%%%%%% 

\subsection{Sedimentation}

In a previous a report, the anti-thixotropic behavior of CB dispersions was interpreted as an "apparent" behavior, resulting from the decrease of the hydrodynamic volume fraction of densifying agglomerates, and their concomitant sedimentation~\cite{Hipp2019}. Following the analysis of~\cite{Hipp2019}, the occurrence of sedimentation can be foreseen using the Shield number defined as the ratio of viscous ($F_v$) to gravitational ($F_g$) forces when considering sedimentation over the cluster size:

\begin{equation}
    S = \frac{F_v}{F_g} = \frac{9\eta_f\dot{\gamma}}{2\xi_c g \Delta \rho}
\end{equation}

\noindent with $g$ the gravitational constant, $\xi_c$ the characteristic cluster size, $\Delta \rho$ the difference of densities between cluster of particles and background fluid, $\eta_f$ the viscosity of the background fluid. 

\begin{figure}[t!]
    \includegraphics[scale=0.53, clip=true, trim=0mm 15mm 0mm 25mm]{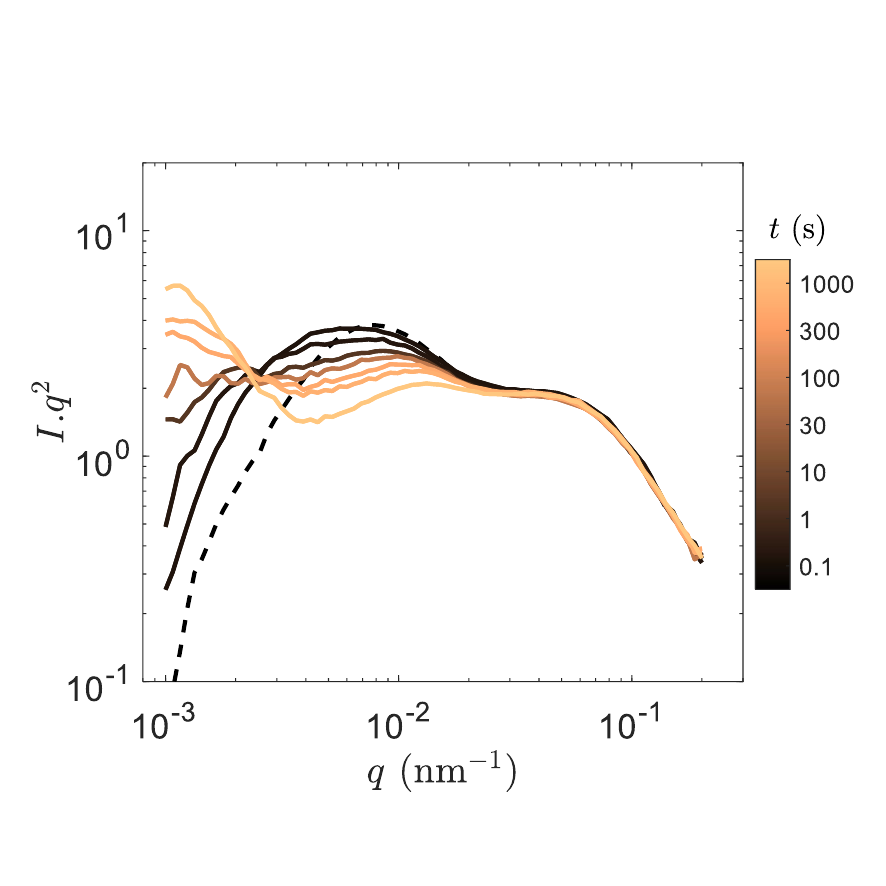}
    \centering
    \caption{Kratky plots $I.q^2$ vs.~$q$ of the scattering profiles acquired at different times after imposing a shear rate of $\dot{\gamma}_0 = 1~\rm s^{-1}$ for the 3.2 \% of CB dispersion. The black dotted curve is the scattering profile of the dispersion during the preshear step at $\dot{\gamma} = 500~\rm s^{-1}$. The intensity corresponding to the form factor of primary CB particles (at $q \approx 7.10^{-2}~\rm nm^{-1}$) remains constant during the anti-thixotropic stress decay, indicating that the volume fraction of CB particles $\phi_{r_0}$ remained constant over time.
    }
    \label{fig:supSedim}
\end{figure}

The cluster density can be calculated from the internal volume fraction of CB particles in a cluster, $\phi_{int} = (\xi_c/r_0)^{(3-d_f)}$, with the cluster radius $\xi_c \approx 1.4~\rm \mu m$, the fractal dimension $d_f = 2.5$ and the CB particle radius $r_0 = 85~\rm nm$~\cite{Bauland2024}. Then, the cluster density is given by $d_c = \phi_{int}.d_{cb} + (1-\phi_{int}).d_{oil} = 1.28.10^3~\rm kg.m^{-3}$. Taking, $\eta_f = 0.25~\rm Pa.s$ and $g = 9.8~\rm m.s^{-2}$, we found $S=1$ for $\dot{\gamma} < 10^{-2}~\rm s^{-1}$. It indicates that sedimentation is not likely to occur over the shear rates investigated. This assumption is supported by the scattered intensity associated with the CB particles length scale, $I(q > 7.10^{-2}~\rm nm^{-1}) = \phi_{r_0}\Delta \rho_{sld}V_{r_0}^2P(q)$, that remains constant over $t = 2\times10^3~\rm s$, which is the longest acquisition time in USAXS (Figure~\ref{fig:supSedim}). This indicates that the volume fraction of CB particles remains constant, ruling out sedimentation effects as an explanation for the initial stress decay with the characteristic time $\tau$. 
%Sedimentation of agglomerates may occur at longer times and could explain the queue relaxation of the stress, observed typically for $t \geq 5.10^3~\rm s$ [Figure~\ref{fig:rheol}(a)]. 

%%%%%%%%%%%%%%%%%%%%%%%%%%%%%%%%%%%%%%%%%%%%%%%%%%%%%%%%%%%%%%%%%%%%%%%%%%%%%%%%%%%%%%
%%%%%%%%%%%%%%%%%%%%%%%%%%%%%%%%%%%%%%%%%%%%%%%%%%%%%%%%%%%%%%%%%%%%%%%%%%%%%%%%%%%%%%

\clearpage 

%\setcitestyle{numerical,square}
%\bibliographystyle{dcu}
%\bibliography{library}
%merlin.mbs aipnum4-1.bst 2010-07-25 4.21a (PWD, AO, DPC) hacked
%Control: key (0)
%Control: author (8) initials jnrlst
%Control: editor formatted (1) identically to author
%Control: production of article title (0) allowed
%Control: page (1) range
%Control: year (1) truncated
%Control: production of eprint (0) enabled
%

\end{document}